%% file: paper_adp_pwmapproach_for_arXiv_v2.tex
\documentclass[journal]{IEEEtran}
\usepackage{cite}

\ifCLASSINFOpdf
  \usepackage[pdftex]{graphicx}
\else
  \usepackage[dvips]{graphicx}
\fi
\usepackage{amsmath}
\usepackage{algpseudocode}

\usepackage{algorithm}
\makeatletter
\newcommand{\removelatexerror}{\let\@latex@error\@gobble}
\makeatother

\usepackage{url}

\usepackage{amssymb}  
\usepackage{etoolbox} 
\usepackage{tikz} 
\usetikzlibrary{fit,shapes.misc,calc}
\input{sec_for_arXiv_v2/00_myCustomCommands}

\input{sec_for_arXiv_v2/00_myCustomCommands_for_ADP}

\newtoggle{doublecolumn}
\begin{document}

\toggletrue{doublecolumn}
%

%
\title{
	Accelerated Point-wise Maximum Approach to Approximate Dynamic Programming
}
%
%
%

\author{Paul~N.~Beuchat,
	Joseph~Warrington,~\IEEEmembership{Member,~IEEE},~
	and~John~Lygeros,~\IEEEmembership{Fellow,~IEEE}
	\thanks{
		This work was supported by the European Research Council under the project OCAL.%
	}
	\thanks{
		$^1$ The authors are with the Automatic Control Laboratory at ETH Z\"{u}rich, Switzerland, {\tt\footnotesize \{beuchatp,warrington,jlygeros\}@ethz.ch}%
	}
}
\maketitle

%

\input{sec_for_arXiv_v2/01_abstract}

\input{sec_for_arXiv_v2/02_intro}

\input{sec_for_arXiv_v2/03_dp}

\input{sec_for_arXiv_v2/04_adp}

\input{sec_for_arXiv_v2/05_numerical}

\input{sec_for_arXiv_v2/06_conclusion}

%
\appendices
%

\input{sec_for_arXiv_v2/11_appendix_pwm_ineq_reformulation}

\input{sec_for_arXiv_v2/11_appendix_inner_problem_proofs}

\input{sec_for_arXiv_v2/11_appendix_formulation_for_solver}






\bibliographystyle{IEEEtran}
\bibliography{paper_adp_pwmapproach_for_arXiv_v2.bib}

\input{sec_for_arXiv_v2/99_biography}

\end{document}

%% file: sec_for_arXiv_v2/00_myCustomCommands.tex
\usepackage{xcolor,calc}

\newtheorem{theorem}{Theorem}[section]
\newtheorem{lemma}[theorem]{Lemma}

\newtheorem{assumption}[theorem]{Assumption}

\newcommand{\mbb}{\mathbb}

\newcommand{\mrm}{\mathrm}
\newcommand{\mcal}{\mathcal}

\newcommand{\tran}{\intercal}

\newcommand{\rdim}[1]{\mathbb{R}^{#1}}

\newcommand{\sdim}[1]{\mathbb{S}^{#1}}

\newcommand{\ndim}[1]{\mathbb{N}^{#1}}

\newcommand{\expval}[2]{\mathbb{E}_{#1}\left[#2\right]}
\newcommand{\trace}[1]{\mathrm{tr}\left(#1\right)}

\newcommand{\diag}[1]{\mathrm{diag}\left(#1\right)}

\newcommand{\setdef}[2]{\left\{{#1} \middle| #2 \right\}}

\newcommand{\matb}[1]{\begin{bmatrix}#1\end{bmatrix}}

\newcommand{\subjto}{\text{s.t.}}

\newcommand{\argmin}[1]{\underset{#1}{\arg\min}\,\,}
\newcommand{\argmax}[1]{\underset{#1}{\arg\max}\,\,}

\newcommand{\arrr}{\rightarrow}
\newcommand{\arrl}{\leftarrow}

\newcommand{\arrR}{\Rightarrow}
\newcommand{\arrL}{\Leftarrow}
\newcommand{\arrLR}{\Leftrightarrow}

\newcommand{\intd}[1]{\mathrm{d}#1}

\newcommand{\regsubdiff}{\hat{\partial}}
\newcommand{\gensubdiff}{\partial}

\newcommand{\gennormalcone} [1]{N_{#1}}

\newcommand{\constraintset}{\mathcal{C}}

\definecolor{light-grey}{gray}{0.80}
\definecolor{mid-grey}{gray}{0.50}
\definecolor{dark-grey}{gray}{0.25}

\definecolor{gray10}{gray}{0.10}
\definecolor{gray20}{gray}{0.20}
\definecolor{gray30}{gray}{0.30}
\definecolor{gray40}{gray}{0.40}
\definecolor{gray50}{gray}{0.50}
\definecolor{gray60}{gray}{0.60}
\definecolor{gray70}{gray}{0.70}
\definecolor{gray80}{gray}{0.80}
\definecolor{gray90}{gray}{0.90}
\definecolorset{rgb}{}{}{myred,     0.7922,0.0000,0.1255} 
\definecolorset{rgb}{}{}{myltred,   0.9569,0.6471,0.5098} 
\definecolorset{rgb}{}{}{myblue,    0.1216,0.4706,0.7059} 
\definecolorset{rgb}{}{}{myltblue,  0.6510,0.8078,0.8902} 
\definecolorset{rgb}{}{}{mygreen,   0.2000,0.6275,0.1725} 
\definecolorset{rgb}{}{}{myltgreen, 0.6980,0.8745,0.5412} 
\definecolorset{rgb}{}{}{mypurple,  0.5961,0.3059,0.6392} 
\definecolorset{rgb}{}{}{myltpurple,0.8706,0.7961,0.8941} 
\definecolorset{rgb}{}{}{myorange,  0.9882,0.5529,0.3490} 
\definecolorset{rgb}{}{}{myyellow,  0.9961,0.8784,0.5647} 

\definecolorset{rgb}{}{}{matlabred,    0.8941,0.1020,0.1098} 
\definecolorset{rgb}{}{}{matlabblue,   0.2157,0.4941,0.7216} 
\definecolorset{rgb}{}{}{matlabpurple, 0.5961,0.3059,0.6392} 
\definecolorset{rgb}{}{}{matlaborange, 0.8500,0.3250,0.0980} 



%% file: sec_for_arXiv_v2/00_myCustomCommands_for_ADP.tex
\newcommand{\dynamics}{g}

\newcommand{\spaceX}{\mathcal{X}}
\newcommand{\spaceU}{\mathcal{U}}
\newcommand{\spaceXU}{\mathcal{X} \! \times \! \mathcal{U}}

\newcommand{\spaceXi}{\Xi}
\newcommand{\xinX}{x\in\mathcal{X}}

\newcommand{\uinU}{u\in\mathcal{U}}

\newcommand{\xinXcompact}{x\!\in\!\mathcal{X}}

\newcommand{\uinUcompact}{u\!\in\!\mathcal{U}}

\newcommand{\funcSpaceX}{\mathcal{F}\!\left(\mathcal{X}\right)}
\newcommand{\approxFuncSpaceX}{\hat{\mathcal{F}}\!\left(\mathcal{X}\right)}

\newcommand{\funcSpaceXU}{\mathcal{F}\!\left(\mathcal{X}\!\times\!\mathcal{U}\right)}
\newcommand{\approxFuncSpaceXU}{\hat{\mathcal{F}}\!\left(\mathcal{X}\!\times\!\mathcal{U}\right)}

\newcommand{\disfac}{\gamma}
\newcommand{\bellmanOp}[2][]{\mathcal{T}_{#1}^{#2}}

\newcommand{\Vopt}{V^\ast}
\newcommand{\Vbasis}{\phi}
\newcommand{\alphaVbasis}{\alpha^\tran \hspace{-0.02cm} \Vbasis}

\newcommand{\Vhat}{\hat{V}}

\newcommand{\Vpwm}{V_{\mathrm{PWM}}}
\newcommand{\EVof}{\mathbb{V}}

\newcommand{\setepiV}{\mathcal{S}}

\newcommand{\BIsufficient}{BI}

\newcommand{\Vobjfixed}{\bar{V}_{\mathrm{obj}}}
\newcommand{\Vconfixed}{\bar{V}_{\mathrm{con}}}

\newcommand{\Aobjfixed}{\mathcal{A}_{\mathrm{obj}}}
\newcommand{\Aconfixed}{\mathcal{A}_{\mathrm{con}}}
\newcommand{\alphaobjfixed}{\bar{\alpha}_{\mathrm{obj}}}
\newcommand{\alphaconfixed}{\bar{\alpha}_{\mathrm{con}}}

\newcommand{\pwmobjfuncsymbol}{f_{\mathrm{pwm}}}

\newcommand{\policyopt}{\pi^{\ast}}


%% file: sec_for_arXiv_v2/01_abstract.tex

\begin{abstract}

We describe an approximate dynamic programming approach to compute lower bounds on the optimal value function for a discrete time, continuous space, infinite horizon setting. The approach iteratively constructs a family of lower bounding approximate value functions by using the so-called Bellman inequality. The novelty of our approach is that, at each iteration, we aim to compute an approximate value function that maximizes the point-wise maximum taken with the family of approximate value functions computed thus far. This leads to a non-convex objective, and we propose a gradient ascent algorithm to find stationary points by solving a sequence of convex optimization problems. We provide convergence guarantees for our algorithm and an interpretation for how the gradient computation relates to the state relevance weighting parameter appearing in related approximate dynamic programming approaches. We demonstrate through numerical examples that, when compared to existing approaches, the algorithm we propose computes tighter sub-optimality bounds with less computation time.

\end{abstract}

%% file: sec_for_arXiv_v2/02_intro.tex

\section{Introduction} \label{sec:intro}

Many important challenges in science and engineering can be cast in the problem formulation of infinite horizon stochastic optimal control (SOC), from climate control of a building \cite{jones_2012_buildings_journal} to control of a cell population \cite{khammash_2016_cell_integral_feedback}.
The goal of such problems is to find a state feedback policy that minimizes an infinite-horizon discounted cost function.
For a general SOC problem instance, the solution, i.e., the optimal policy, is typically characterized by the theory of dynamic programming (DP) \cite{bellman_1952_theory,bertsekas_1996_stochopt_book,hernandez_2012_discreteTimeMCP}.
%
However, in all but a few special cases, solving the SOC directly or applying the DP theory is intractable due to the so-called \emph{curse of dimensionality}.
%
As such, an extensive body of literature has proposed approximation techniques for computing sub-optimal solutions to SOC problems, ranging from model-free and simulation-based algorithms \cite{sutton_2018_rlbook}, to model-based approaches \cite{rawlings_1999_MPCtextbook,goulart_2006_robust_policies,nemirovski_2006_chance_constrained_programs,bertsekas_2005_fromADPtoMPC}. Any technique based on DP theory falls in the category of Approximate DP (ADP), see \cite{2004_adp_handbook,powell_2011_ADP_book,bertsekas_2012_DP_book_vol2} for an overview.
%
Although the optimal policy is intractable to compute, techniques have been developed to bound the sub-optimality of an approximate policy. These type of bounds provide the designer with valuable information about the potential benefit of synthesizing and evaluating alternative policies. In this paper we consider and propose approaches that provide sub-optimality bounds based on the so-called \emph{Linear Programming (LP) approach to ADP} \cite{schweitzer_1985_originalMDP}.

%
The LP approach to ADP provides sub-optimality bounds by computing approximations that are lower bounds of the so-called \emph{value function}, i.e., the solution of the Bellman equation for DP \cite{bellman_1952_theory}.
%
The LP approach parameterizes approximate value functions as a linear combination of fixed basis functions and uses the so-called \emph{Bellman inequality} to restrict consideration to only those linear combinations that are point-wise lower bounds of the value function. To compute an approximate value function the designer specifies the regions of the state space that are of interest via the \emph{state relevance weighting} function, and then solves an optimization problem to find a linear combination of basis functions that maximizes the integral with respect to this weighting.
%
The LP approach was first proposed for finite state and input spaces in \cite{depenoux_1960_original_lp_approach}, and equipped with theoretical guarantees in \cite{vanRoy_2003_lpapproach}. The authors of \cite{vanRoy_2003_lpapproach} also provide a discussion on the importance and difficulty of choosing the state relevance weighting to give the best performance and lower bound. An iterated version of the Bellman inequality was proposed in \cite{boyd_2014_iterated_bellman_inequality} and used to compute tighter lower bounds, however, the topic of choosing the state relevance weighting is not addressed. The subsequent works \cite{boyd_2011_minmax_adp,boyd_2013_iteratedApproxValueFunctions} avoid the need for a state relevance weighting by focusing on the design of policies rather than providing tighter lower bounds.
%
In \cite{summers_ADPwithSOS} the authors use sum-of-squares programming techniques to compute high-order polynomial approximate value functions using the iterated Bellman inequality. The use of high-order polynomials would reduce the difficulty of choosing the state relevance weighting, however, the optimization problem to solve becomes formidable.
%

%
Given a family of lower bounding approximate value functions, computed via the LP approach to ADP, taking a point-wise maximum over the family will yield the same or better approximation of the value function. However, to the best of our knowledge, no algorithm exists that explicitly aims to maximize this point-wise maximum combination; attempting to do this directly leads to a non-convex optimization problem.
%
The benefit of a point-wise maximum combination is empirically demonstrated in \cite{boyd_2014_iterated_bellman_inequality} for a simple example, with the set of state relevance weighting parameters hand-picked using problem-specific insight.
%
In our previous work \cite{beuchat_2017_pwm_at_CDC}, we proposed a problem formulation with the point-wise maximum combination used in the Bellman inequality. The formulation was used to develop an iterative algorithm for computing lower bounding approximate value functions, however, the quality of the approximation, comparable with that of \cite{boyd_2014_iterated_bellman_inequality}, still relies on the designer choosing a sequence of state relevance weightings.
%
The algorithm proposed in \cite{hohmann_2018_moment_ddp} also uses the point-wise maximum combination in the Bellman inequality, and the authors propose an algorithm that computes the sequence of state relevance weightings based on simulating the evolution of the system in a so-called \emph{forward pass}. They consider a finite horizon setting and it is not clear how to extend the algorithm to an infinite horizon setting.
%
A variety of other ADP algorithms compute lower bounds using theoretical tools different from the Bellman inequality, for example \cite{pereira_1991_ddp,rantzer_2006_relaxing_dp,bart_2013_performance_bounds}, each with its advantages and disadvantages, and none of which are similar to the algorithms we propose.

In this paper, we propose a formulation that explicitly aims to maximize a point-wise maximum of lower bounding approximate value functions, and we use this to develop novel algorithms for computing sub-optimality bounds.
In particular, the contributions of the paper are:
\begin{itemize}
	\item We introduce the point-wise maximum formulation of DP and prove that it is equivalent to the LP approach under standard assumptions for SOC problems.
	
	\item We propose a gradient ascent algorithm for finding approximate solutions to the point-wise maximum formulation, and prove that it converges to stationary points. We provide an interpretation of the gradient ascent steps as an algorithmic choice of the state relevance weighting for the non-convex problem of maximizing the point-wise maximum objective.
	
	\item We propose an algorithm for computing initial conditions for the gradient ascent and prove that it converges in finite iterations for any tolerance. This algorithm is required because, for non-convex problems, the quality of a gradient ascent solution is influenced by the initial condition.
\end{itemize}
In support of the contributions, we provide numerical results to demonstrate the sub-optimality bounds achieved and computation time required.
Section~\ref{sec:dp} presents the point-wise maximum DP formulation.
Section~\ref{sec:adp} introduces the approximation methods and our proposed algorithms.
Section~\ref{sec:numerical} demonstrates the performance through numerical examples.

%% file: sec_for_arXiv_v2/03_dp.tex

\section{Dynamic Programming (DP) Formulation} \label{sec:dp}

\subsection{Stochastic Optimal Control Formulation and Assumptions} \label{sec:dp:soc}

%
We consider discrete time, infinite horizon, discounted cost, stochastic optimal control problems over continuous state and action spaces.
%
%
The state of the system at time $t$ is denoted by $\smash{x_t \!\in\! \spaceX \subseteq \rdim{n_x}}$. The system state is influenced by the control decisions $\smash{u_t \!\in\! \spaceU \subseteq \rdim{n_u}}$, and by the stochastic exogenous disturbance $\smash{\xi_t \!\in\! \spaceXi \subseteq \rdim{n_\xi}}$. In this setting, the states evolves according to the function $\smash{\dynamics:\spaceX \!\times\! \spaceU \!\times\! \spaceXi \rightarrow \spaceX}$ as $\smash{x_{t+1} = \dynamics\left( x_t , u_t , \xi_t \right)}$, incurring the stage cost $\disfac^t \, l\left(x_t,u_t\right)$ at each time step, where $\smash{\disfac \!\in\! \left[0,1\right)}$ is the discount factor.
%
By $\Pi$ we denote the set of all feasible deterministic Markov policies, defined as $\smash{\setdef{\pi(\cdot)}{\pi(x) \!\in\! \spaceU ,\, \forall \xinXcompact}}$.
%
The goal is to find a policy $\smash{x_t \mapsto \pi\left(x_t\right)}$ that minimizes the  cumulative cost over an infinite horizon, with initial condition $\xinXcompact$,
	\begin{equation} \label{eq:soc_general}
		\begin{aligned}
			\Vopt(x) \,:=\,
				\inf\limits_{\pi\in\Pi} \hspace{0.15cm}&
				\mbb{E} \left[\, \sum\nolimits_{t=0}^{\infty} \, \disfac^t \, l(x_t,u_t) \,\right]
			\\
			\subjto \hspace{0.15cm}& x_{t+1} \,=\, \dynamics\left( x_t , u_t , \xi_t \right)
				\,,\hspace{0.2cm}
				\forall t \geq 0
				\,,
			\\
			& u_t = \pi(x_t)
				\,,\hspace{1.7cm}
				\forall t \geq 0
				\,,
			\\
			& x_t \in \mcal{X}
				\,,\hspace{0.2cm}
				u_t \in \mcal{U}
				\,,\hspace{0.85cm}
				\forall t \geq 0
				\,,
			\\
			& x_0 = x
				\,.
		\end{aligned}
	\end{equation}
The function $\smash{\Vopt : \spaceX \arrr \rdim{}}$ is the value function that represents the optimal cost-to-go from any state of the system if the optimal control policy is played.

%
To ensure that the problem is well posed we work in that same setting as \cite[\S 6.3]{hernandez_2012_discreteTimeMCP}, specifically under \cite[Assumption~4.2.1(a)]{hernandez_2012_discreteTimeMCP} that the stage cost is lower semi-continuous, non-negative, and inf-compact, and also under \cite[Assumptions 4.2.1(b), 4.2.2]{hernandez_2012_discreteTimeMCP}.
%
The assumptions ensure that from the class of time-varying stochastic policies, the minimum is attained by a stationary deterministic policy, see \cite[Theorem 4.2.3]{hernandez_2012_discreteTimeMCP}.
%
Finally, $\funcSpaceXU$ and $\funcSpaceX$ are defined as the vector spaces of bounded, real-valued, Borel-measurable functions on $\spaceXU$ and $\spaceX$ respectively, where \cite[Definition 6.3.2, 6.3.4]{hernandez_2012_discreteTimeMCP} provides the definitions of boundedness.

\subsection{Linear Programming (LP) Formulation of DP} \label{sec:dp:lp}

Solving the stochastic optimal control problem is equivalent to finding $\Vopt$ as the solution of the Bellman equation \cite{bellman_1952_theory},
	\begin{equation} \label{eq:bellman}
		\Vopt(x) = \underbrace{
				\inf\limits_{\uinU}
				\,\,
				\overbrace{
					\Big\{\,
					l(x,u) \,+\, \disfac \, \expval{}{\Vopt\left( \dynamics(x,u,\xi) \right)}
					\,\Big\}
					}^{\left(\bellmanOp[u]{}\Vopt\right)(x,u)}
			}_{\left(\bellmanOp{}\Vopt\right)(x)}
			\,,\hspace{0.1cm}
			\forall \, \xinXcompact
			\,.
	\end{equation}
$\bellmanOp{}$ is known as the Bellman operator, and the $\bellmanOp[u]{}$ operator represents the cost of making decision $u$ now and then playing optimally from the next time step forward. The optimal policy can be defined using $\Vopt$ by,
	\begin{equation} \label{eq:greedy_policy}
		\policyopt(x) \,=\,
			\left\{\,
			\argmin{\uinU} \hspace{0.1cm} l(x,u) \,+\, \disfac \, \expval{}{\Vopt\left( \dynamics(x,u,\xi) \right)}
			\,\right\}
			\,.
	\end{equation}
The existence of a $\Vopt$ and $\policyopt$ that are Borel measurable and attain the infimum is ensured by \cite[Assumptions 4.2.1(a), 4.2.1(b), 4.2.2]{hernandez_2012_discreteTimeMCP}.
%
%
If $\nu(\cdot)$ is a finite measure on $\spaceX$ that assigns positive mass to all open subsets of $\spaceX$, then it can be shown that the solutions of the following linear program,
	\begin{subequations} \label{eq:lp_for_V}
		\begin{align}
			\max_{V} \hspace{0.15cm}&
				\int\nolimits_{\spaceX} \, V(x) \, \nu(\intd{x})
				\label{eq:lp_for_V:objective}
			\\
			\subjto \hspace{0.15cm}&
				V \in \funcSpaceX
			\\
			& V(x) \,\leq\, \left( \bellmanOp[u]{} V \right) (x,u)
				\,, \hspace{0.3cm}
				\forall \, \xinX ,\, \uinU
				\label{eq:lp_for_V:bellman_ineq}
		\end{align}
	\end{subequations}
satisfy \eqref{eq:bellman} for $\nu$-almost all ($\nu$-a.a.) $\smash{\xinXcompact}$, see \cite[\S 6.3]{hernandez_2012_discreteTimeMCP}. Constraint \eqref{eq:lp_for_V:bellman_ineq} is referred to as the \emph{Bellman Inequality}.
%
%
A key feature of the LP formulation is that any choice of $\nu(\cdot)$ that places mass over the whole state space $\spaceX$ leads \eqref{eq:lp_for_V} to recover a solution of the stochastic optimal control problem.

\subsection{Point-wise Maximum Formulation of DP} \label{sec:dp:pwm}

Following \cite{beuchat_2017_pwm_at_CDC}, we introduce additional decision variables and use a point-wise maximum of value functions in the objective and the Bellman inequality constraint,
	\begin{subequations} \label{eq:pwm_nlp_for_V}
		\begin{align}
			\max_{V_1 , \dots , V_J} \hspace{0.15cm}&
				\int\nolimits_{\spaceX} \, \Vpwm(x) \, \nu(\intd{x})
				\label{eq:pwm_nlp_for_V:objecitve}
			\\
			\subjto \hspace{0.15cm}&
				V_j \in \funcSpaceX
				\,,\hspace{0.3cm}
				j=1,\dots,J
			\\
			& \Vpwm(x) \leq \left( \bellmanOp[u]{} \Vpwm \right) (x,u)
				, \hspace{0.14cm}
				\forall \, \xinXcompact ,\, \uinUcompact
				\label{eq:pwm_nlp_for_V:pwm_ineq}
			\\
			& \Vpwm(x) = \max\limits_{j=1,\dots,J} \, V_j(x)
				, \hspace{0.60cm}
				\forall \, \xinXcompact
		\end{align}
	\end{subequations}
where $\smash{J\!\in\!\ndim{}}$ specifies the number of value function decision variables.
%
We refer to problem \eqref{eq:pwm_nlp_for_V} as the \emph{point-wise maximum formulation}, and the key difference from \cite{beuchat_2017_pwm_at_CDC} is the use of the point-wise maximum $\Vpwm$ in the objective \eqref{eq:pwm_nlp_for_V:objecitve}.
%
The following lemma establish some important properties of \eqref{eq:pwm_nlp_for_V}.

\vspace{0.1cm}

\begin{lemma} \label{thm:pwm_equivalence}
	Problems \eqref{eq:lp_for_V} and \eqref{eq:pwm_nlp_for_V} are equivalent in the sense that there exist mappings between the feasible solutions and the optimal solutions of the two problems. Moreover, objective \eqref{eq:pwm_nlp_for_V:objecitve} is jointly convex in the decision variables $V_j$, $\smash{j\!=\!1,\dots,J}$.
\end{lemma}

\begin{IEEEproof}
	Under the assumptions and definitions of Section~\ref{sec:dp:soc}, one can easily see there is a mapping between feasible solutions since the space $\funcSpaceX$ is closed under the maximum operation, i.e., $V_j \!=\! V$ for all $\xinXcompact$, $\smash{j\!=\!1,\dots,J}$ in one direction, and $V \!=\! \Vpwm$ for all $\xinXcompact$ in the other direction. This gives equivalent objective value by construction, and thus $\int \Vopt d\nu$ is the optimal value for both \eqref{eq:lp_for_V} and \eqref{eq:pwm_nlp_for_V}.
	
	The function $\Vpwm$ is convex in $V_j$, $\smash{j\!=\! 1,\dots,J}$ by definition of the $\max$ function over a finite number of elements. Thus \eqref{eq:pwm_nlp_for_V:objecitve} is convex  as integration is a linear operation, see for example \cite[Lemma 2.1]{bertsekas_1973_stochopt}.
\end{IEEEproof}

\vspace{0.1cm}

The reason for the point-wise maximum formulation becomes apparent when we consider approximating the solution to \eqref{eq:lp_for_V} and \eqref{eq:pwm_nlp_for_V} by restricting the space of the decision variables to subspaces of $\funcSpaceX$. To gain some initial insight, observe that the Bellman inequality constraint \eqref{eq:pwm_nlp_for_V:pwm_ineq} implies that feasible decisions for \eqref{eq:pwm_nlp_for_V} will be point-wise under-estimators of $\Vopt$. Thus a point-wise maximum is a natural way to combine a family of feasible but sub-optimal decisions.

Computing a solution of problem \eqref{eq:pwm_nlp_for_V} poses the following difficulties
\begin{enumerate}
	\renewcommand{\labelenumi}{(D\theenumi)}
	\item $\funcSpaceX$ is an infinite dimensional space;
		\label{difficulty:F}
	\item Objective \eqref{eq:pwm_nlp_for_V:objecitve} involves a multidimensional integral over $\spaceX$;
		\label{difficulty:intobj}
	\item The $\bellmanOp[u]{}$-operator involves a multidimensional dimensional integral over $\spaceXi$;
		\label{difficulty:Tu}
	\item Constraint \eqref{eq:pwm_nlp_for_V:pwm_ineq} involves an infinite number of constraints;
		\label{difficulty:infcon}
	\item Constraint \eqref{eq:pwm_nlp_for_V:pwm_ineq} is non-convex in the decision variables;
		\label{difficulty:nccon}
	\item The objective \eqref{eq:pwm_nlp_for_V:objecitve} involves the maximization of a convex function;
		\label{difficulty:ncobj}
\end{enumerate}
Difficulties (D1-D4) apply also to problem \eqref{eq:lp_for_V} and a variety of approaches have been proposed to address them, see for example \cite{swaroop_2010_pwcValueFunc,vanroy_2004_sampDP,boyd_2012_quadraticADP,sutter_2014_ADPsamp,summers_2017_inf_dim_LPapproach}. In Section \ref{sec:adp} we take inspiration from previous approaches to propose an approximation algorithm that additionally overcomes difficulties (D5-D6).

%% file: sec_for_arXiv_v2/04_adp.tex

\section{Approximate Dynamic Programming (ADP)} \label{sec:adp}

This section proposes an algorithm for computing an approximate value function that is feasible for problem \eqref{eq:pwm_nlp_for_V} at every iteration and analyzes its convergence.

\subsection{Approaches Adopted for Difficulties (D\ref{difficulty:F}), (D\ref{difficulty:Tu}) and (D\ref{difficulty:infcon})} \label{sec:adp:existing}

To overcome difficulty (D1), as suggested in \cite{schweitzer_1985_originalMDP}, we restrict the value functions candidates to the span of a finite family of Borel-measurable basis functions $\smash{\Vbasis_k \!:\! \spaceX \!\arrr\! \rdim{}}$, $\smash{k\!=\!1,\dots,K}$. We parameterize the restricted function space as,
	\begin{equation} \label{eq:approx_func_spaces}
		\approxFuncSpaceX = \setdef{ \alpha^\tran \, \Vbasis(x)}{ \alpha\in\rdim{K} }
		,\hspace{0.20cm}
		\text{with}
		\;
		\Vbasis(x) = \matb{\Vbasis_1(x) \\ \vdots \\ \Vbasis_K(x) }
		\hspace{-0.00cm}.
	\end{equation}
The benefit of this parameterization is that it is linear in the $\alpha$ parameter.
%
Each approximate value function $\Vhat_j$ is parameterized by its own vector that we denote $\alpha_j$, i.e., $\smash{\Vhat_j(x) \!=\! \alpha_j^\tran \, \Vbasis(x)}$ for all $\xinXcompact$.
%
For the numerical examples in Section~\ref{sec:numerical} we use the space of polynomial functions up to a certain degree by choosing the $\Vbasis_k$ to be each of the monomials up to that degree.

To overcome difficulty (D\ref{difficulty:Tu}) we first use Jensen's inequality to switch the order of expectation and maximisation in the $\bellmanOp[u]{}\Vpwm$ term, thus providing a sufficient condition for constraint \eqref{eq:pwm_nlp_for_V:pwm_ineq}. We then require that for each basis function $\smash{\expval{}{\Vbasis\left( g(x,u,\xi)\right)}}$ has an analytic expression.
%
In the case of polynomial basis functions and polynomial dynamics, this requires knowledge of the moments of the distribution of $\xi$ up to the maximum degree of $\xi$ in $\Vbasis\left( g(x,u,\xi)\right)$. If the required moments are not analytically available, then the Monte Carlo sampling can be used to approximate them, and, as the distribution is stationary, this only needs to be computed once.

To overcome difficulty (D\ref{difficulty:infcon}), a variety of convex sufficient conditions techniques are proposed in the literature for approximating \eqref{eq:pwm_nlp_for_V:pwm_ineq} with a finite number of constraints, for example \cite{boyd_2014_iterated_bellman_inequality, summers_ADPwithSOS,lasserre_2009_soc_via_occupation_measures,nikos_2013_ADPforReachability}. The applicable reformulation depends on the problem data and basis functions, and the algorithm we propose in the sequel applies for all such convex inner approximations.
%
For example, when all problem data is polynomial and polynomial basis functions are used, then constraint \eqref{eq:pwm_nlp_for_V:pwm_ineq} can be inner approximated using the sum-of-squares (SOS) S-procedure \cite{summers_ADPwithSOS}.

\subsection{Proposed Approach for Difficulties (D\ref{difficulty:intobj}), (D\ref{difficulty:nccon}) and (D\ref{difficulty:ncobj})} \label{sec:adp:pwm_objective}

%
The inclusion of the point-wise maximum value function in the objective \eqref{eq:pwm_nlp_for_V:objecitve} is pivotal in the algorithm we propose, however, it precludes the use of previous approaches for evaluating the integral in the objective.
%
To overcome difficulty (D\ref{difficulty:intobj}) we replace $\nu$ by a finitely supported measure denoted $c$. Specifically, we choose $c$ as a finite sum of $N_c$ Dirac pulses located at  $\smash{ \{x_{c,i}\}_{i=1}^{N_c} \subset \spaceX}$.
%
This violates the hypothesis for equivalence between \eqref{eq:bellman} and \eqref{eq:lp_for_V}, but reduces the multidimensional integral in \eqref{eq:pwm_nlp_for_V:objecitve} to a sum over the locations of the Dirac pulses.

For clarity of presentation, we consider now an auxiliary problem that highlights our proposed approach for overcoming difficulties (D\ref{difficulty:nccon}) and (D\ref{difficulty:ncobj}).
%
%
We consider two families of functions defined by two finite sets $\smash{\Aobjfixed, \Aconfixed \!\subset\! \rdim{K}}$. The first family, parameterized by $\smash{\alphaobjfixed \!\in\! \Aobjfixed}$ is used in a point-wise maximum objective, while the second, parameterized by $\smash{\alphaconfixed \!\in\! \Aconfixed}$, is used in a point-wise maximum constraint. We then define,
\begin{subequations}
	\begin{align}
	\Vobjfixed(x) = \max_{\alphaobjfixed\in\Aobjfixed} \hspace{0.2cm}
	\alphaobjfixed^\tran \phi(x)
	\,,\hspace{0.2cm} \forall \, \xinXcompact
	,
	\\
	\Vconfixed(x) = \max_{\alphaconfixed\in\Aconfixed} \hspace{0.2cm}
	\alphaconfixed^\tran \phi(x)
	\,,\hspace{0.2cm} \forall \, \xinXcompact
	,
	\end{align}
\end{subequations}
and assume that $\Aobjfixed$ and $\Aconfixed$ have been selected such that $\smash{\Vobjfixed \!\leq\! \Vopt}$ and $\smash{\Vconfixed \!\leq\! \Vopt}$ for all $\smash{\xinXcompact}$.
%
%
Additionally, we introduce $\smash{\pwmobjfuncsymbol : \rdim{K} \arrr \rdim{}}$ as the point-wise maximum objective function when adding an additional function $\smash{\alpha^\tran \Vbasis(x) \!\in\! \approxFuncSpaceX}$ to the function $\Vobjfixed$, i.e.,
%
\begin{equation}
	\pwmobjfuncsymbol\left(\alpha \right) = \frac{1}{N_c} \sum\limits_{i=1}^{N_c} \, \max\left\{ \alphaVbasis(x_{c,i}) , \Vobjfixed(x_{c,i}) \right\}
		\,,
\end{equation}
where $x_{c,i}$ are the points selected to overcome difficulty (D\ref{difficulty:intobj}).
%
The auxiliary problem for maximizing $\pwmobjfuncsymbol$ in the presence of $\Vobjfixed$ and $\Vconfixed$ is,
%
\begin{subequations} \label{eq:canonical_inner_problem}
	\begin{align}
	\max_{\alpha \in \rdim{K}} \hspace{0.15cm}&
	\pwmobjfuncsymbol\left(\alpha \right)
	\label{eq:canonical_inner_problem:objective}
	\\
	\subjto \hspace{0.15cm}&
	\alphaVbasis(x) \leq \left( \bellmanOp[u]{} \Vconfixed \right) (x,u)
	\,, \hspace{0.2cm}
	\forall \, \xinXcompact ,\, \uinUcompact
	\,.
	\label{eq:canonical_inner_problem:pwm_ineq}
	\end{align}
\end{subequations}

With $\Vconfixed$ as fixed parameters in \eqref{eq:canonical_inner_problem} the constraint is convex in the decision variable $\alpha$, thus overcoming difficulty (D\ref{difficulty:nccon}). Moreover, if $\Vconfixed$ satisfies the Bellman inequality, then \eqref{eq:canonical_inner_problem:pwm_ineq} implies that $\max\left\{\alphaVbasis,\Vconfixed\right\}$ also satisfies the Bellman inequality. The steps to show convexity of constraint \eqref{eq:canonical_inner_problem:pwm_ineq}, first presented in \cite{beuchat_2017_pwm_at_CDC}, are provided in Appendix \ref{app:pwm_ineq_reformulation} for completeness.
%
In Section~\ref{sec:adp:pwm_algorithm} we present the proposed algorithm for iteratively adding elements to $\Aconfixed$ in a greedy fashion, where the difference compared to \cite{beuchat_2017_pwm_at_CDC} is the choice of objective weighting.
%
To simplify the presentation, we introduce the notation,
\begin{equation} \nonumber
	\alpha \in BI\left(\Aconfixed\right) \subseteq \rdim{K}
	\hspace{0.2cm} \arrR \hspace{0.2cm}
	\alpha \text{ satisfies \eqref{eq:canonical_inner_problem:pwm_ineq}}
	\,,
\end{equation}
to represent the convexified Bellman inequality constraint \eqref{eq:canonical_inner_problem:pwm_ineq}.

Despite the convexified constraint, \eqref{eq:canonical_inner_problem} is still a non-convex problem due to (D\ref{difficulty:ncobj}).
%
In general, problem  \eqref{eq:canonical_inner_problem} will have multiple distinct local maxima and stationary points.
%
The convexity of the objective means that given element of the sub-differential, constructed at a particular point in the decision variable space $\alpha$, it parameterizes a hyperplane that is a point-wise lower-bound on the objective function.
%
Thus we propose to iteratively maximize along sub-gradient directions to overcome difficulty (D\ref{difficulty:ncobj}), and in Section~\ref{sec:adp:gradient_algorithm} we introduce the algorithm and its convergence properties.

\newcommand{\algoneidx}{k}
%
\begin{figure}
	\removelatexerror
	%
	\begin{algorithm}[H]
		\caption{Find points satisfying necessary optimality conditions of problem \eqref{eq:canonical_inner_problem} with $c$ as a sum of Dirac pulses}
		\label{alg:inner_problem}
		\begin{algorithmic}[1]
			%
			\Procedure{InnerProblem}{ $\smash{\alpha^{(0)}}$, $\Aobjfixed$, $\Aconfixed$ , $\epsilon$ }
			%
			\State $\algoneidx\gets 0$
			%
			\Repeat
			%
			\State $d^{(\algoneidx)} \gets$ an element from $\gensubdiff^+ \pwmobjfuncsymbol\left( \alpha^{(\algoneidx)} \right)$
			\label{alg:inner_problem:subdiff}
			%
			\If{$\left( d^{(\algoneidx)} = 0 \right)$ }
			\label{alg:inner_problem:zero_subdiff}
			%
			\State $\alpha^{(\algoneidx+1)} \gets \alpha^{(\algoneidx)}$
			\label{alg:inner_problem:zero_subdiff:break}
			%
			\Else
			\State $\alpha^{(\algoneidx+1)} \gets$ $\smash{\argmax{} \!\! \left\{ \alpha^\tran \, d^{(\algoneidx)} \,;\, \alpha \!\in\! \BIsufficient(\Aconfixed) \right\}}$
			\label{alg:inner_problem:lp}
			%
			\EndIf
			%
			\State $\algoneidx\gets \algoneidx+1$
			%
			\Until{$\Big( \pwmobjfuncsymbol\left(\alpha^{(\algoneidx)} \right) - \pwmobjfuncsymbol\left(\alpha^{(\algoneidx-1)} \right) \Big) < \epsilon$.}
			\label{alg:inner_problem:equal_objective}
			%
			\State \textbf{return} $\alpha^{(\algoneidx)}$
			%
			\EndProcedure
		\end{algorithmic}
	\end{algorithm}
\end{figure}

\subsection{First-order method for the point-wise maximum objective}\label{sec:adp:gradient_algorithm}

%
We propose Algorithm \ref{alg:inner_problem} to improve $\pwmobjfuncsymbol$ from a given feasible initial condition $\alpha^{(0)}$ using only first-order information of the objective function.
%
The objective $\pwmobjfuncsymbol$ is in general non-smooth as it is a maximum of functions, thus in line \ref{alg:inner_problem:subdiff} we use the upper sub-differential for selecting gradient ascent directions, denoted as $\gensubdiff^+ \pwmobjfuncsymbol$ and defined in Appendix \ref{app:differentiability_definitions}; an upper sub-differential is considered because \eqref{eq:canonical_inner_problem} is a maximization problem.
%
Given a non-zero element from the upper sub-differential, i.e., an upper sub-gradient, in line \ref{alg:inner_problem:lp} we update the decision variable by maximizing along the sub-gradient direction within the feasible region.
The algorithm terminates when the change in objective value between two subsequent iterations is less than a pre-specified tolerance.

To compute an element from the upper sub-differential of $\pwmobjfuncsymbol$, to be used in line \ref{alg:inner_problem:subdiff} of Algorithm~\ref{alg:inner_problem}, we introduce the following assumption on the basis functions.

\vspace{0.1cm}

\begin{assumption} \label{ass:basisfunctions}
	The basis functions in the set $\smash{ \left\{ \Vbasis_k \right\}_{k=1}^{K} }$ are continuous for all $\xinXcompact$ and include the constant function.
\end{assumption}

\vspace{0.1cm}

%
Without loss of generality we take $\smash{ \Vbasis_1(x) \,=\, 1 }$ for all $\xinXcompact$.
%
For a general choice of basis functions $\Vbasis_k$ it is difficult to characterize the upper sub-differential set at non-smooth points.
%
Instead, we work with a particular element from the upper sub-differential of $\pwmobjfuncsymbol$ that is readily computable at $\alpha$ under Assumption \ref{ass:basisfunctions},
%
\begin{equation} \label{eq:pwmobj_subdiff}
	\gensubdiff^+ \pwmobjfuncsymbol\left( \alpha \right) \ni
		\frac{1}{N_c} \sum\limits_{i=1}^{N_c}
		\begin{cases}
			\Vbasis(x_{c,i}) & \text{if}\, \alphaVbasis(x_{c,i}) \geq \Vobjfixed(x_{c,i})
			\\
			0 & \text{if}\, \alphaVbasis(x_{c,i}) < \Vobjfixed(x_{c,i})
		\end{cases}
		,
\end{equation}
where the term inside the sum is an element from the upper sub-differential of $\smash{\max\left\{ \alphaVbasis(x) , \Vobjfixed(x) \right\}}$.
%
%
Given the element $\smash{\bar{d} \in \gensubdiff^+ \pwmobjfuncsymbol}\left( \bar{\alpha} \right)$ computed as per \eqref{eq:pwmobj_subdiff} at a point $\smash{\bar{\alpha}\!\in\!\rdim{K}}$, the hyperplane
%
$\smash{\left( \alpha - \bar{\alpha} \right)^\tran \bar{d} + f\left( \bar{\alpha} \right)}$
is a supporting hyperplane of the convex function $\pwmobjfuncsymbol$. However, we note that for maximization of a convex function not all supporting hyperplanes are in the upper sub-differential.
%
%
In the proof of Theorem \ref{thm:inner_problem_algorithm_convergence} we show that \eqref{eq:pwmobj_subdiff} is indeed an element of the upper sub-differential.

Algorithm \ref{alg:inner_problem} can be seen as a method that iteratively adjusts the objective of line~\ref{alg:inner_problem:lp} along sub-gradient directions of problem~\eqref{eq:canonical_inner_problem}.
%
%
To ensure that an element from the argmax can always be computed in line~\ref{alg:inner_problem:lp}, we introduce the following assumption on the choice of basis functions and inner approximation set.

\vspace{0.1cm}

\begin{assumption} \label{ass:lp_attainment}
	The basis function set $\smash{ \left\{ \Vbasis_k \right\}_{k=1}^{K} }$, Bellman inequality inner approximation set $\BIsufficient(\Aconfixed)$, and problem data are such that the following optimization problem,
		\begin{equation} \label{eq:adp:lp_approach:dirac_objective}
			\max\limits_{\alpha\in\rdim{K}} \left\{\,
				\alphaVbasis(x_{c,i})
				\hspace{0.1cm};\hspace{0.1cm}
				\alpha \in \BIsufficient(\Aconfixed)
				\,\right\}
		\end{equation}
	attains its maximum for all $\smash{ \{x_{c,i}\}_{i=1}^{N_c} }$.
\end{assumption}

\vspace{0.1cm}

This assumption is not overly restrictive as it can be ensured for any general problem instance by placing an upper bound on a norm of $\alpha$, see \cite{sutter_2018_infinite_to_finite} for example. For a particular problem instance the assumption can be verified by, for example, showing the existence of a strictly feasible point in the dual of \eqref{eq:adp:lp_approach:dirac_objective} \cite[Theorem 3.1]{boyd_1996_sdp} \cite[Corollary 30.5.2]{rockafellar_2015_convexanalysis}.

\vspace{0.2cm}

\begin{theorem} \label{thm:inner_problem_algorithm_convergence}
	Under Assumptions~\ref{ass:basisfunctions} and~\ref{ass:lp_attainment}, for any initial condition $\smash{\alpha^{(0)} \!\in\! \BIsufficient(\Aconfixed)}$ and any $\smash{\epsilon \!>\! 0}$, Algorithm \ref{alg:inner_problem} generates a non-decreasing sequence $\smash{ \pwmobjfuncsymbol\left(\alpha^{(\algoneidx)}\right)}$ and terminates after a finite number of iterations.
	%
	With $\smash{\epsilon \!=\! 0}$, the sequence $\smash{ \pwmobjfuncsymbol\left(\alpha^{(\algoneidx)}\right)}$, converges to a finite value, and the sequences $\smash{ \alpha^{(\algoneidx)} }$, $\smash{ d^{(\algoneidx)} }$, satisfy the following necessary optimality condition for $\smash{ \max\nolimits_{\alpha\in\rdim{K}} \left\{ \pwmobjfuncsymbol(\alpha) \hspace{0.01cm};\hspace{0.05cm} \alpha \!\in\! \BIsufficient(\Aconfixed) \right\} }$ in the limit,
		\begin{equation} \nonumber
			\lim\limits_{\algoneidx \arrr \infty}\left(\,
			\min\limits_{\alpha \in \BIsufficient(\Aconfixed)} \, \left(\alpha-\alpha^{(\algoneidx)}\right)^\tran d^{(\algoneidx)}
			\,\right)
			\,=\, 0
			\,.
		\end{equation}
\end{theorem}

\vspace{0.1cm}

\begin{IEEEproof}
	see Appendix \ref{app:inner_problem_algorithm_convergence_proof}.
\end{IEEEproof}

\vspace{0.2cm}

Theorem \ref{thm:inner_problem_algorithm_convergence} guarantees that if the initial condition $\alpha^{(0)}$ strictly improves on $\pwmobjfuncsymbol(0)$, then $\pwmobjfuncsymbol(\alpha^{(i)})$ returned also strictly improves on $\pwmobjfuncsymbol(0)$. The convergence in finite iterations ensures that the algorithm is practical to implement, and the limiting behaviour suggests that in the best case the $\alpha^{(i)}$ returned could be close to a local maxima. Although Theorem \ref{thm:inner_problem_algorithm_convergence} provides no insight into the rate of convergence, the numerical examples in Section \ref{sec:numerical} demonstrate that significant improvement in $\pwmobjfuncsymbol$ can be achieved with only a handful of iterations.

%
Algorithm \ref{alg:inner_problem} is a so-called \emph{Minorize Maximize} algorithm for maximizing a convex function, and we now contrast with generic algorithms that exist in the literature for this same purpose.
%
In the case where $\pwmobjfuncsymbol$ is differentiable, then line \ref{alg:inner_problem:subdiff} of Algorithm \ref{alg:inner_problem} becomes $\smash{d^{(\algoneidx)} \arrl \nabla \pwmobjfuncsymbol( \alpha^{(\algoneidx)})}$ and is a special-case of the so-called \emph{convex-concave procedure} introduced in \cite{yuille_2003_cccp}, and for which convergence guarantees are given in \cite[Theorem 4]{lanckriet_2009_cccp_convergence}.
%
Algorithms applicable for non-smooth problems like \eqref{eq:canonical_inner_problem} are presented together with convergence guarantees in \cite[Theorem 3]{pham_1997_dca} and \cite[Proposition 1]{koulik_2018_cccp_nonsmooth}. Applying the algorithm from \cite{pham_1997_dca} or \cite{koulik_2018_cccp_nonsmooth} to problem \eqref{eq:canonical_inner_problem} would require using the lower sub-differential of $\pwmobjfuncsymbol$ in line \ref{alg:inner_problem:subdiff} of Algorithm \ref{alg:inner_problem}.
%
For a non-smooth convex function the lower sub-differential contains the upper sub-differential, thus allowing more flexibility on line \ref{alg:inner_problem:subdiff} of Algorithm \ref{alg:inner_problem}.
%
However, \cite{pham_1997_dca} and \cite{koulik_2018_cccp_nonsmooth} use the lower sub-differential also for defining necessary optimality conditions. This means that, compared to Algorithm \ref{alg:inner_problem}, the algorithms from \cite{pham_1997_dca} and \cite{koulik_2018_cccp_nonsmooth} may have additional points in their convergence set that are not local maxima of the non-smooth convex maximization problem.

\newcommand{\algtwoidx}{m}
\begin{figure}
	\removelatexerror
	%
	\begin{algorithm}[H]
		\caption{Maximise the value of $\int \Vobjfixed dc$}
		\label{alg:outer_problem}
		\begin{algorithmic}[1]
			\Procedure{OuterProblem}{}
			
			\State \textbf{Select} $\Aobjfixed$ , $\Aconfixed$ , $\smash{ \{x_{c,i}\}_{i=1}^{N_c} }$ according to \textsection\ref{sec:adp:discussion}
			
			\State \textbf{Select} $\epsilon_{\mathrm{IN}} ,\, \epsilon_{\mathrm{OUT}} \,<\, 0$
			
			\State $\algtwoidx\gets 0$
			
			\Repeat{}
			
			\State $f^{(\algtwoidx)} \gets$ $\frac{1}{N_c} \sum\limits_{i=1}^{N_c} \left(\, \max\limits_{\alphaobjfixed\in\Aobjfixed} \hspace{0.1cm} \alphaobjfixed^\tran \phi(x_{c,i}) \,\right)$
			\vspace{0.1cm}
			\label{alg:outer_problem:start_objective}
			
			\ForAll{ $\{x_{c,i}\}_{i=1}^{N_c}$ }
			\label{alg:outer_problem:start_for}
			
			\State $\alpha^{(0)} \gets$ $\smash{\argmax{} \!\! \left\{ \alphaVbasis(x_{c,i}) ; \alpha \!\in\! \BIsufficient(\Aconfixed) \right\}}$
			\label{alg:outer_problem:generate}
			
			\State $\tilde{\alpha} \gets$ \textsc{InnerProblem}$\left(\smash{\alpha^{(0)}}, \Vobjfixed, \Vconfixed, \epsilon_{\mathrm{IN}} \right)$
			\label{alg:outer_problem:refine}
			
			\State $\Aobjfixed \gets$ $\tilde{\alpha} \,\cup\,\Aobjfixed$
			
			\State $\Aconfixed \gets$ $\tilde{\alpha} \,\cup\,\Aconfixed$
			
			\EndFor
			\label{alg:outer_problem:end_for}
			
			\State $\algtwoidx\gets \algtwoidx+1$
			
			\vspace{0.1cm}
			\State $f^{(\algtwoidx)} \gets$ $\frac{1}{N_c} \sum\limits_{i=1}^{N_c} \left(\, \max\limits_{\alphaobjfixed\in\Aobjfixed} \hspace{0.1cm} \alphaobjfixed^\tran \phi(x_{c,i}) \,\right)$
			\vspace{0.1cm}
			\label{alg:outer_problem:end_objective}
			
			\Until{$\left( f^{(\algtwoidx)} - f^{(\algtwoidx-1)} \right) < \epsilon_{\mathrm{OUT}}$,}
			\label{alg:outer_problem:equal_objective}
			\State \textbf{return} $\Aobjfixed$ , $\Aconfixed$
			\EndProcedure
		\end{algorithmic}
	\end{algorithm}
\end{figure}

\subsection{Point-wise Maximum ADP Algorithm} \label{sec:adp:pwm_algorithm}

In this section we propose Algorithm \ref{alg:outer_problem}, which iteratively updates the value function estimates used in the objective and constraints of problem \eqref{eq:canonical_inner_problem}, i.e., $\Vobjfixed$ and $\Vconfixed$.
%
At each iteration of lines \ref{alg:outer_problem:start_for}--\ref{alg:outer_problem:end_for}, a candidate approximate value function $\alpha^{(0)}$ is generated by solving \eqref{eq:adp:lp_approach:dirac_objective} with $x_{c,i}$ as one of the Dirac pulse locations from $c(\cdot)$. Algorithm \ref{alg:inner_problem} refines this candidate before it is added to the collections $\Aobjfixed$ and $\Aconfixed$. This process of generating, refining, and adding is repeated for all $x_{c,i}$, $\smash{i=1,\dots,N_c}$. The algorithm terminates when the improvement in $\Vobjfixed$ is below some pre-specified threshold.
%
The following theorem formalises the convergence properties of Algorithm \ref{alg:outer_problem}.

\vspace{0.2cm}

\begin{theorem} \label{thm:outer_problem_algorithm_convergence}
	For any sets  $\Aobjfixed$ and $\Aconfixed$ such that $\Vobjfixed$ and $\Vconfixed$ are point-wise under-estimators of $\Vopt$, and for any $\smash{ \epsilon_{\mathrm{IN}},\, \epsilon_{\mathrm{OUT}} > 0 }$, Algorithm \ref{alg:outer_problem} terminates after a finite number of iterations.
\end{theorem}

\vspace{0.1cm}

\begin{IEEEproof}
	%
	By Theorem \ref{thm:inner_problem_algorithm_convergence} we have that line \ref{alg:outer_problem:refine} of Algorithm \ref{alg:outer_problem} terminates after finite iterations for all $\smash{\epsilon_{\mathrm{IN}}\!>\!0}$.
	The sequence $f^{(\algtwoidx)}$ in non-decreasing by definition as a point-wise maximum of functions and because elements are never removed from the set $\Aobjfixed$. The same reasoning as Appendix \ref{app:inner_problem_algorithm_convergence_proof} establishes that $\smash{ \max\nolimits_{\left\{ \alphaobjfixed\in\Aobjfixed \right\}} \hspace{0.1cm} \alphaobjfixed^\tran \phi(x)}$ is bounded above for all $\smash{ \{x_{c,i}\}_{i=1}^{N_c} }$ at all iterations of Algorithm \ref{alg:outer_problem}. Hence $f^{(\algtwoidx)}$ is bounded above and is thus a convergent sequence.
	Therefore, for all $\smash{\epsilon_{\mathrm{OUT}} \!>\! 0}$ there must exist an $\smash{\algtwoidx \!\geq\! 1}$ such that the condition on line \ref{alg:outer_problem:equal_objective} triggers.
\end{IEEEproof}

\vspace{0.2cm}

The convergence of Algorithm \ref{alg:outer_problem} is guaranteed even without the refinement steps of Algorithm \ref{alg:inner_problem}. However, our numerical results in Section~\ref{sec:numerical} show that without refinement convergence tends to be much slower, and that significant improvements are achieved with only a few iterations of Algorithm \ref{alg:inner_problem}. Hence, the gradient-based motivation and theoretical guarantees of Theorem \ref{thm:inner_problem_algorithm_convergence} suggests that performing the refinement steps of Algorithm \ref{alg:inner_problem} is beneficial.

The objective $\alphaVbasis(x_{c,i})$ in line \ref{alg:outer_problem:generate} of Algorithm \ref{alg:outer_problem} is chosen so that the $\alpha^{(0)}$ passed to Algorithm \ref{alg:inner_problem} has a non-zero sub-gradient $d^{(0)}$ (line~\ref{alg:inner_problem:subdiff} of Algorithm~\ref{alg:inner_problem}).  To see this, note that the sub-gradient in \eqref{eq:pwmobj_subdiff} is non-zero if $\alphaVbasis(x_{c,i})$ weakly dominates $\Vobjfixed(x_{c,i})$ for at least one $\smash{i=1,\dots,N_c}$. Thus, by Assumption \ref{ass:lp_attainment}, line \ref{alg:outer_problem:generate} of Algorithm \ref{alg:outer_problem} computes an $\alpha^{(0)}$ that weakly dominates $\Vobjfixed$ at the chosen point $x_{c,i}$ if such a solution exists in the feasible set $\alpha \!\in\! \BIsufficient(\Aconfixed)$.
%
Different objectives for line \ref{alg:outer_problem:generate} of Algorithm \ref{alg:outer_problem} can be considered and still enjoy the convergence guarantee of Theorem \ref{thm:outer_problem_algorithm_convergence}. However, this would introduce a tuning parameter and empirical testing has shown no benefit when hand-tuning the objective.

\subsection{Discussion and extensions} \label{sec:adp:discussion}

Considering the motivating problem \eqref{eq:pwm_nlp_for_V} with objective $\int \Vpwm d\nu$, the obvious choice for $c$ is to draw samples from $\nu$, and to choose $\nu$ as the initial state distribution. However, sampling $c$ in different ways may improve the objective $\int \Vpwm d\nu$, and for this reason $c$ is commonly referred to as the \emph{state relevance weighting} \cite{vanRoy_2003_lpapproach}. The sub-gradient computed by \eqref{eq:pwmobj_subdiff} effectively sub-samples the points from $c$ where the current approximate value function dominates the fixed $\Vobjfixed$. Thus Algorithm \ref{alg:inner_problem} can be seen a method for automatically choosing the state relevance weighting parameter to maximize $\pwmobjfuncsymbol$, i.e., the surrogate for $\int \Vpwm d\nu$.

If the goal is to optimize the on-line performance of the greedy policy, it is again likely that difference choices of $\nu$, and hence different samples for $c$, lead to differing on-line performance. Motivated by the performance bounds provided in \cite{vanRoy_2003_lpapproach}, a reasonable choice is to place Algorithm \ref{alg:outer_problem} inside another iteration that updates $\nu$ as the discounted occupancy measure for the current greedy policy, computed empirically by simulating the system evolution using Monte Carlo sampling.

For real-time applications where the greedy policy must be computed very fast, it is necessary that the cardinality of $\Aobjfixed$ is small, and perhaps even a singleton.
%
For examples with linear dynamics, quadratic stage costs, polytopic spaces, and using the space of quadratics for $\approxFuncSpaceX$, then the greedy policy is a Quadratically Constrained Quadratic Program (QCQP), with the number of quadratic constraints equal to the cardinality of $\Aobjfixed$. In such examples, a low cardinality of $\Aobjfixed$ has clear benefits from an on-line compution perspective.
%
In these cases it is beneficial to run Algorithm~\ref{alg:outer_problem} twice. First, Algorithm~\ref{alg:outer_problem} is run for as long as practical to achieve a good under-estimate of $\Vopt$, with a simple initialization, for example $\smash{\Aobjfixed\!=\!\Aconfixed\!=\!\{0\}}$. Second, Algorithm~\ref{alg:outer_problem} is run for as many iterations as the desired cardinality of $\Aobjfixed$, with $\Aconfixed$ initialized as the under-estimate resulting from the first run.

The approximate value function computed by Algorithm~\ref{alg:outer_problem} can be used off-line to certify the empirical performance of alternative policies that do not use the approximate value function.
In this case Algorithm~\ref{alg:outer_problem} is run for as long as practical, then the chosen policy is simulated from a particular initial state, $\hat{x}$, for a time horizon such that $\gamma^t$ has decayed sufficiently. The approximate value function evaluated at the initial state is a lower bound on $\Vopt(\hat{x})$ and thus provides a bound on the sub-optimality of the policy, and hence indicates the potential benefit of considering further alternatives.

In \cite{beuchat_2017_pwm_at_CDC} the value function decision variable was also included in the right-hand-side of constraint \eqref{eq:canonical_inner_problem:pwm_ineq}, i.e.,
\begin{equation} \nonumber
\alphaVbasis(x) \leq \left( \bellmanOp[u]{} \left( \max\left\{ \alphaVbasis(x) , \Vconfixed(x) \right\} \right) \right) (x,u)
\,,
\end{equation}
for all $\xinXcompact$ and $\uinUcompact$.
This results in a bi-linear term in the constraint, and in that work the authors suggest gridding the multiplier of the bi-linear term. We do not consider this extension in the numerical examples because it adds significant computation time and empirically it provides little or no benefit for the examples considered.

Algorithm \ref{alg:inner_problem} can be extended to fit multiple new lower bounding functions at the same time. To exemplify, consider the case of adding two new lower bounding functions. The non-convex optimization problem then becomes,
\begin{subequations} \nonumber 
	\begin{align}
	\max_{\alpha,\beta \in \rdim{K}} \hspace{0.15cm}&
	\frac{1}{N_c} \sum\limits_{i=1}^{N_c} \, \max\left\{ \alphaVbasis(x_{c,i}) , \beta^\tran \hspace{-0.02cm} \Vbasis(x_{c,i}) , \Vobjfixed(x_{c,i}) \right\}
	\\
	\subjto \hspace{0.15cm}&
	\alpha \!\in\! \BIsufficient(\Aconfixed)
	\,, \hspace{0.2cm}
	\beta \!\in\! \BIsufficient(\Aconfixed)
	\,.
	\end{align}
\end{subequations}
We construct an element from the upper subdifferential in a similar fashion,
\begin{equation} \nonumber
\begin{aligned}
&\, \gensubdiff^+\left(\, \max\left\{ \alphaVbasis(x) , \beta^\tran \hspace{-0.02cm} \Vbasis(x) , \Vobjfixed(x) \right\} \,\right)
\\
=&\, \begin{cases}
\matb{ \Vbasis(x)^\tran , 0 }^\tran  & \text{if}\, \alphaVbasis(x) \geq \max\left\{ \beta^\tran \hspace{-0.02cm} \Vbasis(x) , \Vobjfixed(x) \right\}
\\
\matb{ 0 , \Vbasis(x)^\tran }^\tran  & \text{if}\, \beta^\tran \hspace{-0.02cm} \Vbasis(x) > \max\left\{ \alphaVbasis(x) , \Vobjfixed(x) \right\}
\\
\matb{ 0 , 0 }^\tran & \text{if}\, \Vobjfixed(x) > \max\left\{ \alphaVbasis(x) , \beta^\tran \hspace{-0.02cm} \Vbasis(x) \right\}
\end{cases}
\,.
\end{aligned}
\end{equation}
As the constraints are separable we see that once the subdifferential element is computed, then line~\ref{alg:inner_problem:lp} of Algorithm~\ref{alg:inner_problem} can be solved in parallel for $\alpha$ and $\beta$, differing only in the objective vector.

%% file: sec_for_arXiv_v2/05_numerical.tex

\section{Numerical Examples} \label{sec:numerical}

\begin{figure*}
	\centering
	\begin{tikzpicture}
	\coordinate (FO) at (0.0cm,0.0cm);
	\node[inner sep=0pt,anchor=south west] at ($(FO)+(0.5cm,1.0cm)$){
		\includegraphics[width=8.0cm]
		{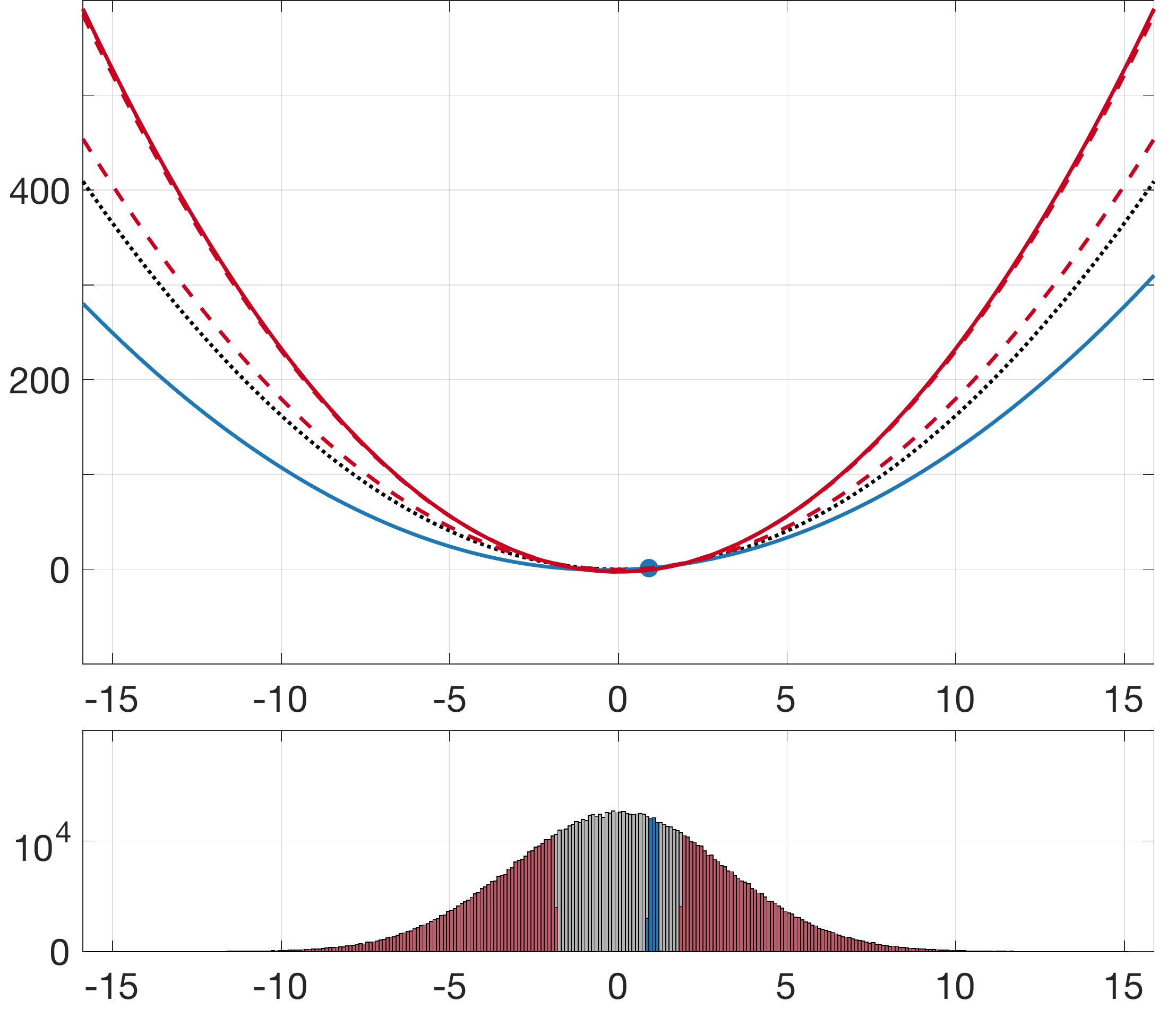}
	};
	\node[inner sep=0pt,anchor=south west] at ($(FO)+(9.5cm,1.0cm)$){
		\includegraphics[width=8.0cm]
		{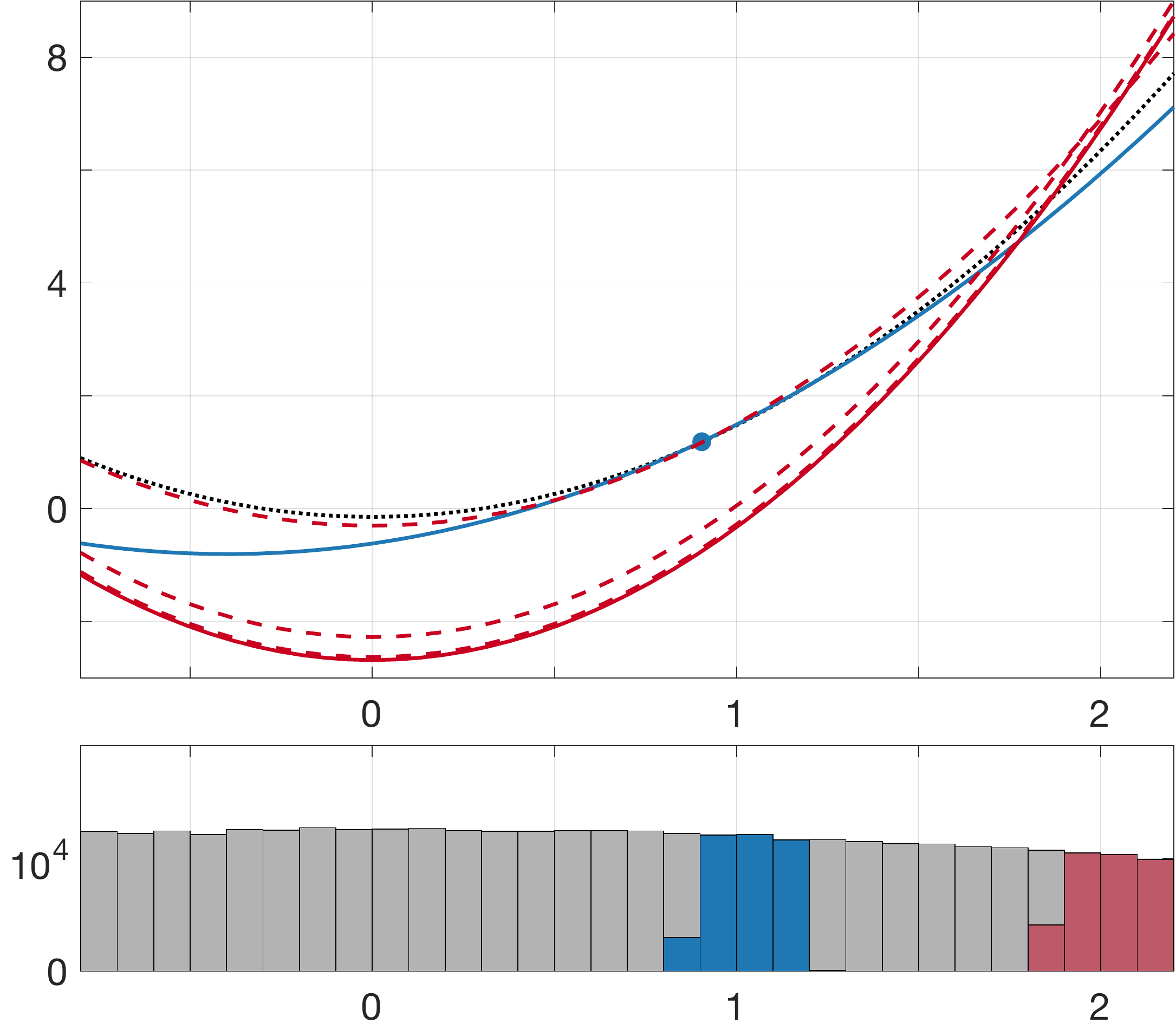}
	};
	%
	%
	\node[align=center , anchor=center , rotate=90] at ($(FO)+(0.2cm,6.0cm)$)
	{{\small Value Function }};
	\node[align=center , anchor=center , rotate=90] at ($(FO)+(9.4cm,6.0cm)$)
	{{\small Value Function }};
	\node[align=center , anchor=center , rotate=0]  at ($(FO)+(4.8cm,0.7cm)$)
	{{\small State Space, $\spaceX = \rdim{}$ }};
	\node[align=center , anchor=center , rotate=90] at ($(FO)+(0.2cm,2.2cm)$)
	{{\small $\nu(x)$ }};
	\node[align=center , anchor=center , rotate=90] at ($(FO)+(9.3cm,2.2cm)$)
	{{\small $\nu(x)$ }};
	\node[align=center , anchor=center , rotate=0]  at ($(FO)+(13.8cm,0.7cm)$)
	{{\small State Space, $\spaceX = \rdim{}$ }};
	%
	%
	\node[align=center , anchor=center , rotate=0] at ($(FO)+( 4.80cm,0.2cm)$)
	{\small{ (a) }};
	\node[align=center , anchor=center , rotate=0] at ($(FO)+(13.80cm,0.2cm)$)
	{\small{ (b) }};
	\end{tikzpicture}
	%
	%
	%
	\caption[Short-hand caption]{
		Providing visual insight for Algorithm \ref{alg:inner_problem} using the 1-dimensional example described in Section \ref{sec:numerical:1d}. Sub-figure (b) is a zoomed view of sub-figure (a).
		%
		On the upper axes, the dotted black line is $\Vobjfixed$ and $\Vconfixed$, the blue dot and blue line are the $x_{c,i}$ and $\alpha^{(0)}$ generated on line \ref{alg:outer_problem:generate} of Algorithm \ref{alg:outer_problem}. The red lines (solid and dashed) are the approximate value functions from the refinement steps of Algorithm \ref{alg:inner_problem}, with the solid line corresponding to the terminal iteration. Algorithm \ref{alg:inner_problem} converged in four steps to a $0.1\%$ relative tolerance on the objective value increase. The lower axes show the $\smash{N_c \!=\! 10^6}$ samples as a histogram, with grey bars showing all samples, blue bars showing where the blue line is greater than $\Vobjfixed$, and red bars showing where the solid red line is greater than $\Vobjfixed$.
	}
	\label{fig:alg_insight}
\end{figure*}

\begin{figure*}
	\centering
	\begin{tikzpicture}
	%
	%
	\coordinate (1DO) at (0.0cm,0.0cm);
	%
	\node[inner sep=0pt,anchor=south west] at ($(1DO)+(0.50cm,3.7cm)$)
	{
		\includegraphics[width=8.0cm]
		{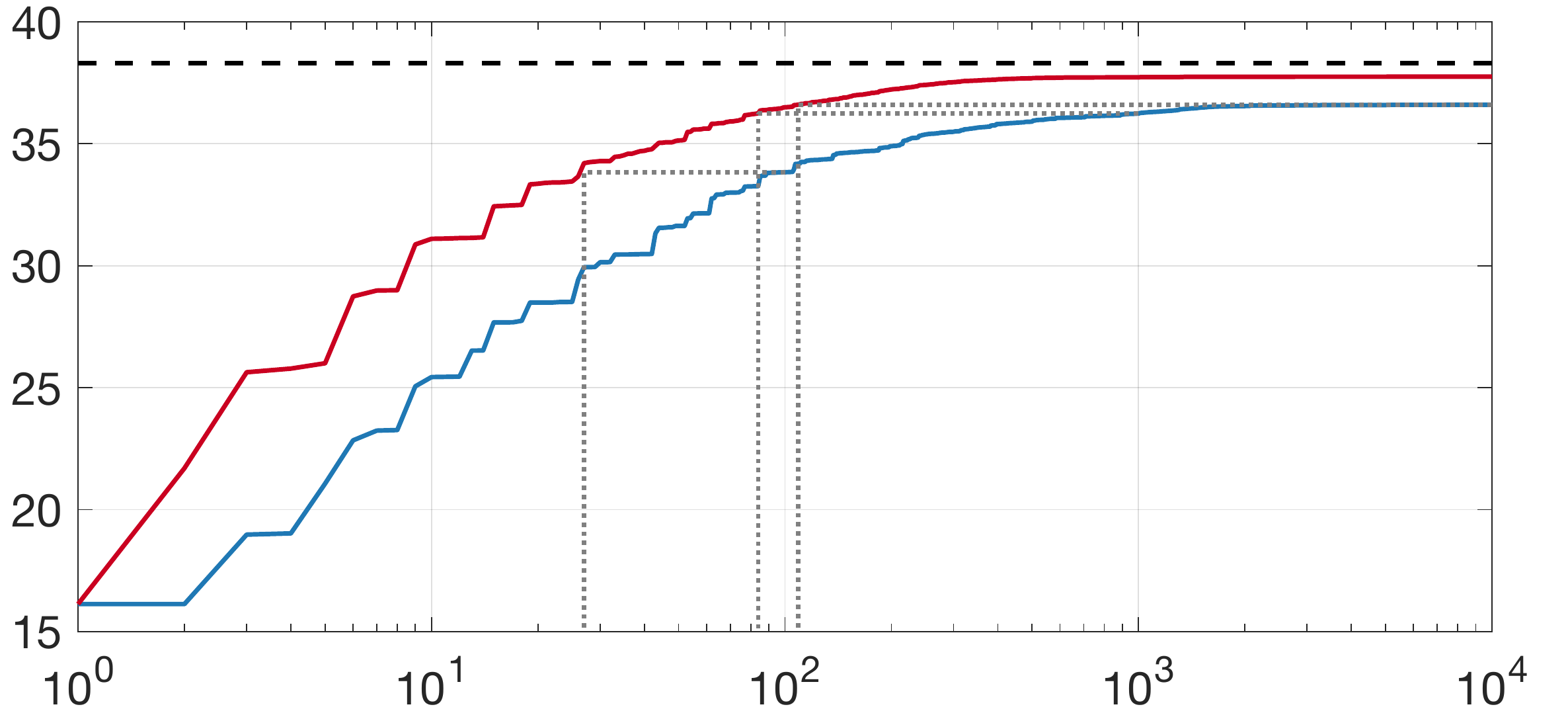}
	};
	\node[inner sep=0pt,anchor=south west] at ($(1DO)+(0.62cm,0.7cm)$)
	{
		\includegraphics[width=7.86cm]
		{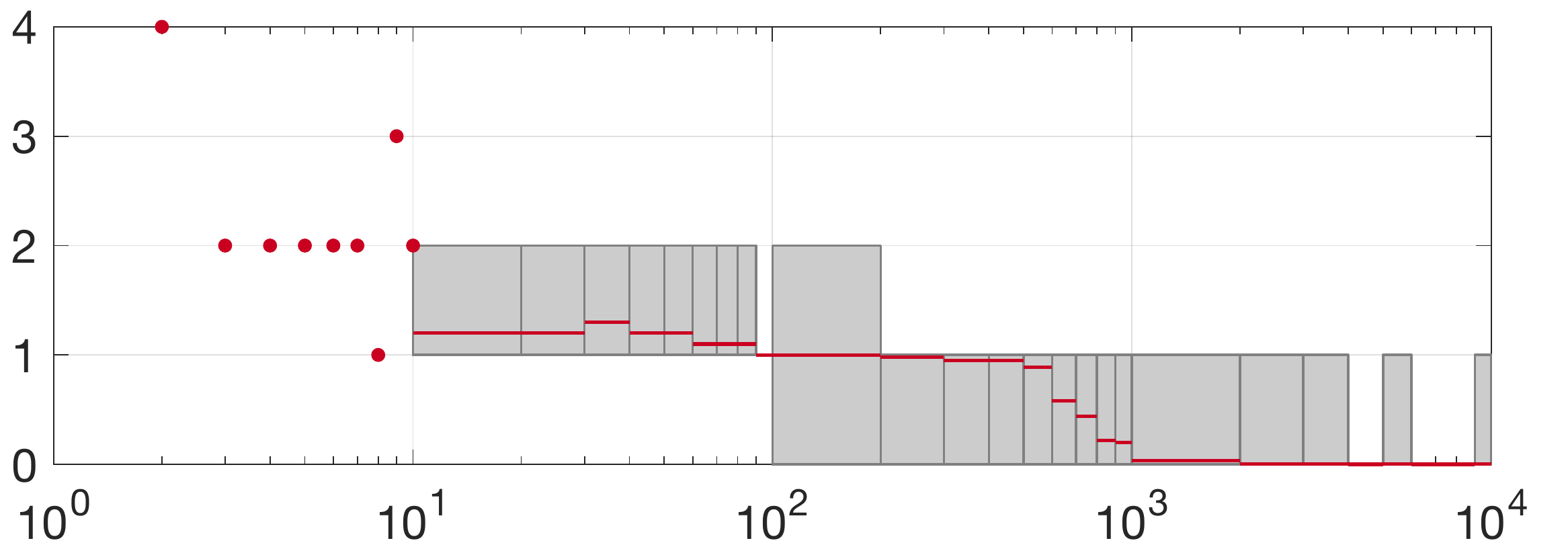}
	};
	%
	%
	\node[align=center , anchor=center , rotate=90] at ($(1DO)+(0.2cm,5.7)$)
	{{\small $\pwmobjfuncsymbol$ }};
	\node[align=center , anchor=center , rotate=90] at ($(1DO)+(0.3cm,2.2)$)
	{{\small \# Iter of Alg. \ref{alg:inner_problem} }};
	\node[align=center , anchor=center , rotate=0]  at ($(1DO)+(4.50cm,0.4)$)
	{{\small \# Iterations of Alg. \ref{alg:outer_problem} }};
	%
	\node[align=center , anchor=center ] at ($(1DO)+(4.25cm,7.6cm)$)
	{\small For $\smash{\spaceX = \rdim{}}$ example of Section~\ref{sec:numerical:1d} };
	%
	%
	%
	%
	%
	\coordinate (10DO) at (9.0cm,0.0cm);
	%
	\node[inner sep=0pt,anchor=south west] at ($(10DO)+(0.50cm,3.7cm)$)
	{
		\includegraphics[width=8.0cm]
		{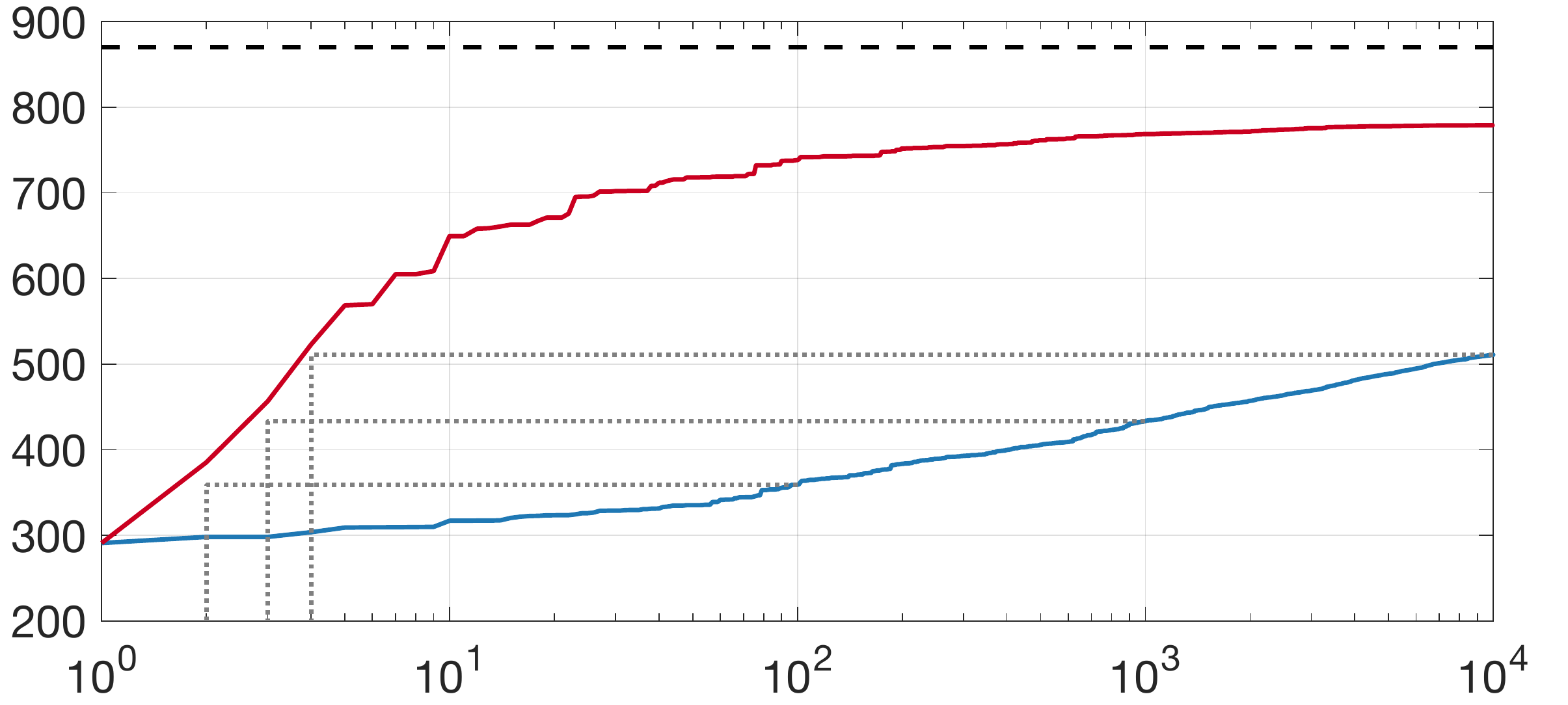}
	};
	\node[inner sep=0pt,anchor=south west] at ($(10DO)+(0.75cm,0.7cm)$)
	{
		\includegraphics[width=7.70cm]
		{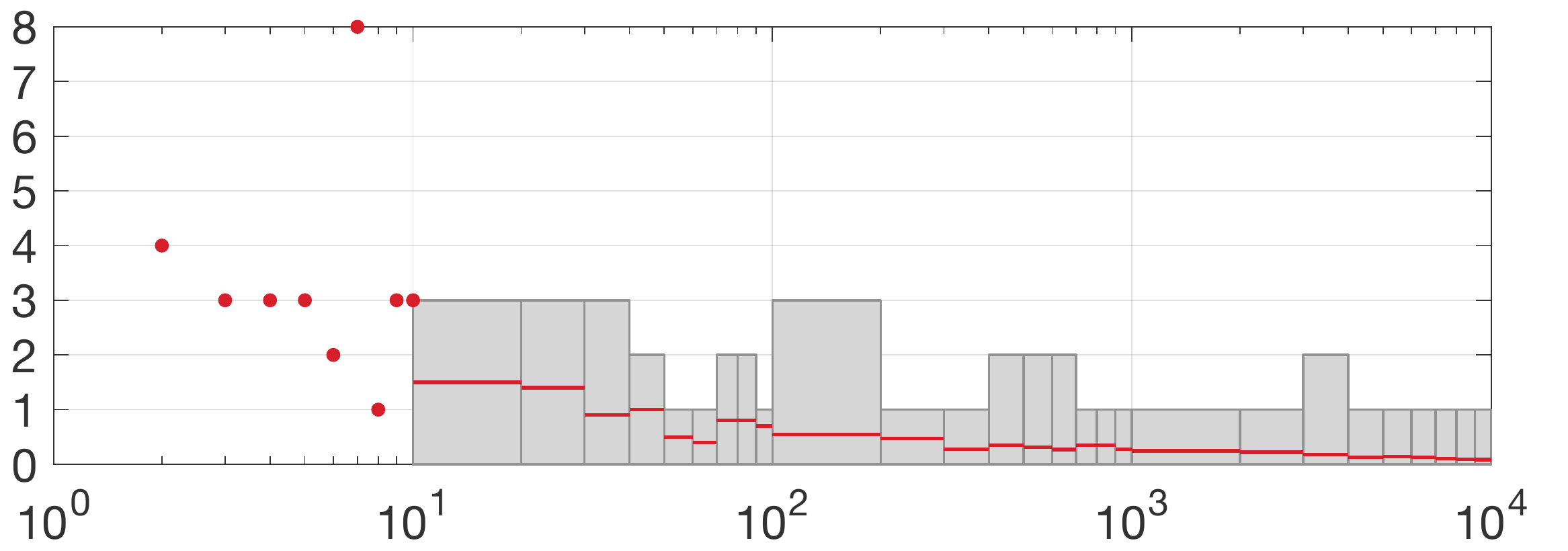}
	};
	\node[align=center , anchor=center , rotate=90] at ($(10DO)+(0.2cm,5.7)$)
	{{\small $\pwmobjfuncsymbol$ }};
	\node[align=center , anchor=center , rotate=90] at ($(10DO)+(0.3cm,2.2)$)
	{{\small \# Iter of Alg. \ref{alg:inner_problem} }};
	\node[align=center , anchor=center , rotate=0]  at ($(10DO)+(4.50cm,0.4)$)
	{{\small \# Iterations of Alg. \ref{alg:outer_problem} }};
	%
	\node[align=center , anchor=center ] at ($(10DO)+(4.25cm,7.6cm)$)
	{\small For $\smash{\spaceX = \rdim{10}}$ example of Section~\ref{sec:numerical:linquad_ND} };
	%
	%
	%
	\coordinate (LO) at (0.5cm,0.0cm);
	\draw[gray70,line width = 0.5pt,fill=white] ($(LO)+(-0.2cm,0.0cm)$) rectangle ($(LO)+(16.6cm,-0.7cm)$);
	\coordinate (LE1) at ($(LO)+(0.0cm,-0.4cm)$);
	\draw[matlabblue,line width = 1.2pt,solid]
	($(LE1)+(0.2cm,0cm)$) -- ($(LE1)+(0.74cm,0cm)$);
	\node[right] at ($(LE1)+(0.85cm,0cm)$) {\color{matlabblue}
		{\small Algorithm \ref{alg:outer_problem} without refinement }};
	\coordinate (LE2) at ($(LE1)+(5.5cm,0.0cm)$);
	\draw[myred,line width = 1.2pt,solid]
	($(LE2)+(0.2cm,0cm)$) -- ($(LE2)+(0.74cm,0cm)$);
	\node[right] at ($(LE2)+(0.85cm,0cm)$) {\color{myred}
		{\small Algorithm \ref{alg:outer_problem} with refinement }};
	\coordinate (LE3) at ($(LE2)+(5.5cm,0.0cm)$);
	\draw[black,line width = 1.2pt,dashed]
	($(LE3)+(0.2cm,0cm)$) -- ($(LE3)+(0.74cm,0cm)$);
	\node[right] at ($(LE3)+(0.85cm,0cm)$) {\color{black}
		{\small Online performance of a policy }};
	\end{tikzpicture}
	%
	%
	%
	\caption[Short-hand caption]{
		Details of the proposed algorithms for: (left) the 1-dimensional example described in Section \ref{sec:numerical:1d}; (right) the 10-dimensional example described in Section \ref{sec:numerical:linquad_ND}.
		%
		The blue lines show the results from running Algorithm \ref{alg:outer_problem} without using Algorithm \ref{alg:inner_problem} to refine the solution at each iteration, while the red lines show the results with refinement.
		The top figures show the point-wise maximum objective integrated with respect to the $N_c$ samples. The bottom figures show number of iterations of Algorithm \ref{alg:inner_problem} until the $\epsilon$-convergence criterion is triggered, i.e., the number of refinement steps performed. To make the bottom figures readable, the results are grouped between the deciles of each order of magnitude, with the horizontal red line showing the average number of iterations, and the grey box spanning the minimum and maximum.
		%
		The online performance of a representative policy is shown by the dotted black line on the top figures. For the 1-dimensional example (left) an LQR policy was used with the input clipped to the constraints, while for the 10-dimensional example (right) a Model Predictive Controller was used with a 10-time-step horizon length and the Riccati equation solution as the terminal cost.
	}
	\label{fig:linquad}
\end{figure*}


In the following numerical example we consider problems with linear dynamics, convex quadratic stage costs, hyper-cube constraints on the input space, and use convex quadratics for the restricted function space $\approxFuncSpaceX$. In Appendix \ref{app:implementation_details} we provide the definition of $\approxFuncSpaceX$, formulate line~\ref{alg:inner_problem:lp} of Algorithm~\ref{alg:inner_problem} so that it can be passed to a standard solver, and verify that Assumption~\ref{ass:basisfunctions} holds.
Assumption~\ref{ass:lp_attainment} was observed to hold empirically, in that the solver returned a finite, optimal solution at each iteration.

\subsection{Algorithm insight on 1-dimensional example} \label{sec:numerical:1d}

To provide visual insight into how Algorithms 1 and 2 adapt the linear combination of basis functions, we use the simple 1-dimension example from \cite{boyd_2014_iterated_bellman_inequality}, i.e., with $\smash{n_x\!=\!n_u \!=\! 1}$.  The dynamics, costs, constraints, and initial state distribution are given by,
\begin{equation} \nonumber
	\begin{aligned}
		&\hspace{0.0cm}
			x_{t+1} = x_t - 0.5 u_t
			\,,\hspace{0.2cm}
			x_0 \sim \mcal{N}(0,10)
			\,,
		\\
		&\hspace{0.0cm}
			|u| \leq 1
			\,,\hspace{0.2cm}
			\gamma = 0.95
			\,,\hspace{0.2cm}
			l(x,u) = x^2 + 0.1 u^2
			\,,
	\end{aligned}
\end{equation}
and we use the space of univariate quadratics as $\approxFuncSpaceX$.
%
We initialise $\Vobjfixed$ and $\Vconfixed$ with the solution of the LP approach using a single Bellman inequality constraint, i.e.,
\begin{equation} \label{eq:LP_single_BI}
	\argmax{\Vhat\in\approxFuncSpaceX} \left\{\,
		\int_{\spaceX} \, \Vhat(x) \, \nu(\intd{x})
		;\hspace{0.1cm}
		\Vhat(x) \leq \bellmanOp{} \Vhat(x),\; \forall \xinX
		\,\right\}
		,
\end{equation}
with $\nu$ chosen as the initial state distribution. The solution of \eqref{eq:LP_single_BI} represents the lower bound proposed in \cite{vanRoy_2003_lpapproach}.
%
In this setting \eqref{eq:LP_single_BI} is a convex semi-definite optimization program, with the objective requiring the first and second moments of $\nu$ and the constraint reformulated as a linear matrix inequality with respect to the quadratic coefficient decision variables, see \cite[\S6]{boyd_2014_iterated_bellman_inequality}.

Choosing $c$ as $\smash{N_c \!=\! 10^6}$ samples from the initial state distribution,
Figure \ref{fig:alg_insight} shows the first iteration of Algorithm \ref{alg:outer_problem}, with sub-figure (b) as zoomed view of sub-figure (a). Sub-figure (a) shows that the candidate approximate value function generated by line \ref{alg:outer_problem:generate} of Algorithm \ref{alg:outer_problem} (blue) is a significantly worse lower bound compared to that resulting from four refinement iterations of Algorithm \ref{alg:inner_problem} (red), when integrated with respect to the $N_c$ samples (grey histogram).

Figure \ref{fig:alg_insight}(b) shows that at the $x_{c,i}$ used on line \ref{alg:outer_problem:generate} of Algorithm \ref{alg:outer_problem} (blue dot), the generated value function (blue line) strictly improves on $\Vobjfixed$ (dotted black line). All of the refinement iterations (red) trade-off a decrease at this $x_{c,i}$ for a significant increase at the other samples. For this example and at this iteration, the significant increase in the point-wise maximum objective value $\pwmobjfuncsymbol$ is gained in regions away from the origin.

%
Figure~\ref{fig:linquad} (top left) shows the lower bound and online performance integrated with respect to the $\smash{N_c \!=\! 10^6}$ samples from the initial state distribution.
%
This shows that 1000 iterations of Algorithm~\ref{alg:outer_problem} combined with the refinement steps of Algorithm~\ref{alg:inner_problem} (red line) allow the clipped-LQR controller to be certified as within 1.5\% of the optimal. Without the refinement steps (blue line) the sub-optimality bound is 4.5\%, even after $10^4$ iterations.
%
Figure~\ref{fig:linquad} (bottom left) shows the number of iterations of Algorithm~\ref{alg:inner_problem} performed at each iteration of Algorithm~\ref{alg:outer_problem}. Together with Figure~\ref{fig:linquad} (top left), this shows that the handful of Algorithm~\ref{alg:inner_problem} iterations improves the bound (red line) with an order of magnitude fewer iterations than Algorithm~\ref{alg:outer_problem} without refinement (blue line).
%
In fact, computations performed for $10^5$ iterations empirically suggest that Algorithm~\ref{alg:outer_problem} without refinement will not give better than a 4.5\% sub-optimality bound within a practical number of iterations, for this example.

\subsection{Higher Dimensional Linear-Quadratic Problems} \label{sec:numerical:linquad_ND}

We consider again the an input constrained linear-quadratic system, this time with dimension $n_x \!=\! 10$, and $n_u \!=\! 3$. The system dynamics take the form,
	\begin{equation} \nonumber
		\begin{aligned}
			x_{t+1} \,=\, A \, x_t \,+\, B_u \, u_t 
				\,,
		\end{aligned}
	\end{equation}
where $A$ and $B_u$ are matrices of compatible size, and the quadratic stage cost is, $\smash{l(x,u) \!=\! x^\tran I_{n_x} x + u^\tran  I_{n_u}  u}$, where $I_n$ denotes an identity matrix of size $n$, and we use discount factor $\smash{\gamma \!=\! 0.99}$.
%
The initial state is normally distributed as $x_0 \!\sim\! \mcal{N}\left( 0 , \Sigma_{\nu} \right)$ with $\Sigma_{\nu} \!=\! 9 I_{n_x}$.
%
The $A$ and $B_u$ matrices are randomly generated, with the $A$ matrix scaled to be marginally stable, i.e., a spectral radius equal to 1. 

%
Figure~\ref{fig:linquad} (top right) shows the lower bound achieved by running Algorithm \ref{alg:outer_problem} without (blue) and with (red) the refinement iterations of Algorithm~\ref{alg:inner_problem}, with the number of refinement iterations shown in Figure~\ref{fig:linquad} (bottom right).
%
To demonstrate the benefit of our proposed Algorithm relative to previous work, the sets $\Aobjfixed$ and $\Aconfixed$ are initialised with the solution of \eqref{eq:LP_single_BI}, with $\nu$ chosen as the initial distribution, i.e., $\mcal{N}\left( 0 , \Sigma_{\nu} \right)$. Thus, the value for iteration 1 of the blue and red lines in Figure~\ref{fig:linquad} (top right), approximately 291, is the lower bound achieved by the method proposed in \cite{vanRoy_2003_lpapproach}.
%
The key feature of the result is that, although Algorithm~\ref{alg:outer_problem} without refinement is guaranteed to converge, the number of iterations required  to reach a reasonable lower bound is significant. Algorithm~\ref{alg:outer_problem} with refinement, on the other hand, achieves a significantly better lower bound with orders of magnitude fewer iterations.

The improvement in the lower bound achieved with the refinement steps of Algorithm \ref{alg:inner_problem} is only meaningful if it significantly tightens the online performance bound for a particular policy. For this example, an MPC policy with a time horizon of $10$ achieves an online performance of $870$, shown by the dotted black line on Figure~\ref{fig:linquad} (top right). Thus the refinement steps of Algorithm~\ref{alg:inner_problem} (red) certify this policy to be within 11\% of the optimal, while the bound without the refinement steps (blue) provides only a 70\% sub-optimality certificate.

\begin{figure}
	\centering
	\begin{tikzpicture}
	%
	%
	\coordinate (10DO) at (0.0cm,0.0cm);
	%
	\node[inner sep=0pt,anchor=south west] at ($(10DO)+(0.50cm,3.7cm)$)
	{
		\includegraphics[width=8.0cm]
		{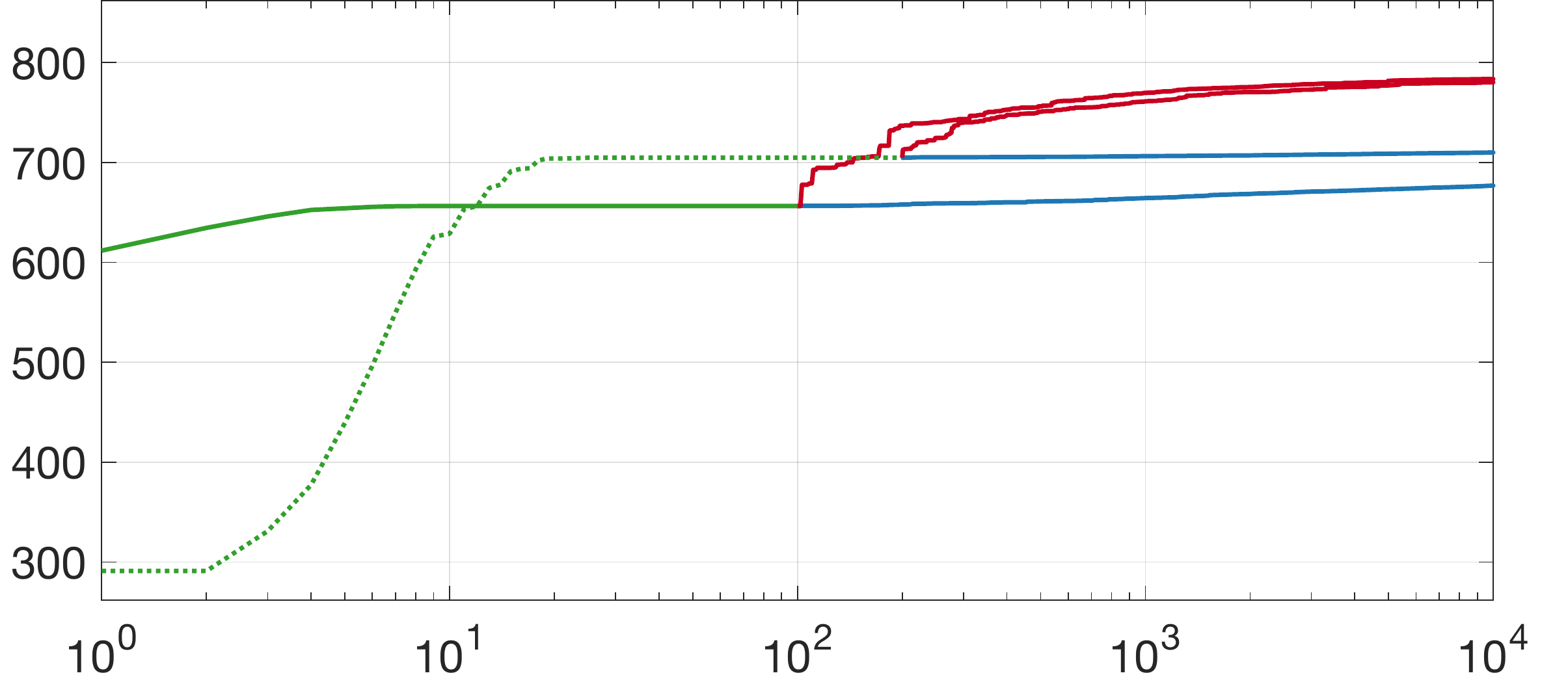}
	};
	%
	\node[inner sep=0pt,anchor=south west] at ($(10DO)+(0.42cm,0.7cm)$)
	{
		\includegraphics[width=8.07cm]
		{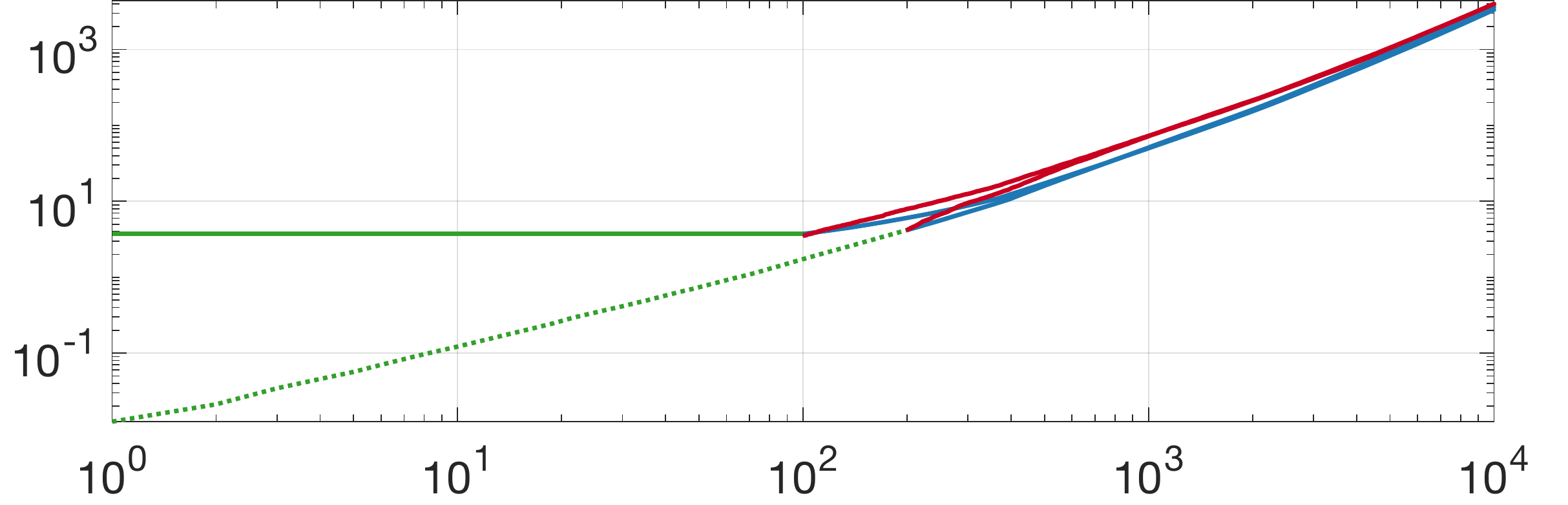}
	};
	%
	%
	\node[align=center , anchor=center , rotate=90] at ($(10DO)+(0.2cm,5.7cm)$)
	{{\small $\pwmobjfuncsymbol$ }};
	\node[align=center , anchor=center , rotate=90] at ($(10DO)+(0.2cm,2.2cm)$)
	{{\small Time [seconds] }};
	\node[align=center , anchor=center , rotate=0]  at ($(10DO)+(4.50cm,0.4cm)$)
	{{\small \# Iterations of Alg. \ref{alg:outer_problem} }};
	%
	%
	\coordinate (LO) at (1.2cm,-0.0cm);
	\draw[gray70,line width = 0.5pt,fill=white] ($(LO)+(-0.2cm,0.0cm)$) rectangle ($(LO)+(6.7cm,-2.0cm)$);
	\coordinate (LE1) at ($(LO)+(0.0cm,-0.4cm)$);
	\draw[mygreen,line width = 1.2pt,solid]
	($(LE1)+(0.2cm,0cm)$) -- ($(LE1)+(0.74cm,0cm)$);
	\node[right] at ($(LE1)+(0.85cm,0cm)$) {\color{mygreen}
		{\small Iterated Bellman Inequality \cite{boyd_2013_iteratedApproxValueFunctions} }};
	\coordinate (LE2) at ($(LE1)+(0.0cm,-0.4cm)$);
	\draw[mygreen,line width = 1.2pt,dotted]
	($(LE2)+(0.2cm,0cm)$) -- ($(LE2)+(0.74cm,0cm)$);
	\node[right] at ($(LE2)+(0.85cm,0cm)$) {\color{mygreen}
		{\small Hand-tuned sequence of $c(x)$ \cite{beuchat_2017_pwm_at_CDC} }};
	\coordinate (LE3) at ($(LE2)+(0.0cm,-0.4cm)$);
	\draw[matlabblue,line width = 1.2pt,solid]
	($(LE3)+(0.2cm,0cm)$) -- ($(LE3)+(0.74cm,0cm)$);
	\node[right] at ($(LE3)+(0.85cm,0cm)$) {\color{matlabblue}
		{\small Algorithm \ref{alg:outer_problem} without refinement }};
	\coordinate (LE4) at ($(LE3)+(0.0cm,-0.4cm)$);
	\draw[myred,line width = 1.2pt,solid]
	($(LE4)+(0.2cm,0cm)$) -- ($(LE4)+(0.74cm,0cm)$);
	\node[right] at ($(LE4)+(0.85cm,0cm)$) {\color{myred}
		{\small Algorithm \ref{alg:outer_problem} with refinement }};
	\end{tikzpicture}
	%
	%
	%
	\caption[Short-hand caption]{
		Comparison with existing methods for computing lower bounds using a point-wise maximum of approximate value functions for the 10-dimensional example of Section~\ref{sec:numerical:linquad_ND}. The top figure is comparable with Figure~\ref{fig:linquad} (top right). The iterated Bellman inequality method \cite{boyd_2013_iteratedApproxValueFunctions} (solid green) is computed with $100$ Bellman inequality iterations, while hand-tuning method \cite{beuchat_2017_pwm_at_CDC} (dotted green) uses a sequence of 20 zero-mean Gaussian distributions in the objective with the variance increasing from $\smash{\Sigma_{\nu} \!=\! 0.1 I_{n_x}}$ to $\smash{\Sigma_{\nu} \!=\! 18 I_{n_x}}$ and repeated 10 times, i.e., 200 iterations total. The bottom figure shows the cumulative computation time for solving line~\ref{alg:inner_problem:lp} of Algorithm~\ref{alg:inner_problem} and line~\ref{alg:outer_problem:generate} of Algorithm~\ref{alg:outer_problem}.
	}
	\label{fig:linquad_with_expanding_cx}
\end{figure}

%
As another point of comparison, we show in Figure~\ref{fig:linquad_with_expanding_cx} (top) the lower bound achieved by two alternative methods (green), and then we use the output of these methods to initialize our proposed algorithm (blue and red lines).
%
For the iterated Bellman inequality method proposed in \cite{boyd_2013_iteratedApproxValueFunctions} (solid green), we use $100$ Bellman inequality iterations, achieving a lower bound of $656$, and we note that more iterations did not improve the bound. This method computes $100$ approximate value functions from the solution of one optimization problem, hence the computation time in Figure~\ref{fig:linquad_with_expanding_cx} (bottom) is constant for the first $100$ iterations, while the lower bound increases because we add the $100$ approximate value functions sequentially to the point-wise maximum approximation.
%
For implementing the hand-tuning method suggested in \cite{beuchat_2017_pwm_at_CDC} (dotted green) we run Algorithm~\ref{alg:outer_problem} without refinement and with the modification that on line \ref{alg:outer_problem:generate} we manually select a different objective at each iteration. We performed this for a range of options and show in Figure~\ref{fig:linquad_with_expanding_cx} the option achieving the tightest bound.
For both of the comparisons, we then use the approximate value functions computed to initialize the sets $\Aobjfixed$ and $\Aconfixed$ and again run Algorithm \ref{alg:outer_problem} without (blue) and with (red) the refinement steps of Algorithm~\ref{alg:inner_problem}.
%
In both cases, only a handful of iterations of Algorithm \ref{alg:inner_problem} are required to achieve the improved lower bound of $783$, an $11\%$ and $19\%$ improvement respectively.

%
The combination of results in Figure~\ref{fig:linquad} (top right) and Figure~\ref{fig:linquad_with_expanding_cx} (top) suggests that the lower bound achieved by Algorithm \ref{alg:outer_problem} with refinement (red lines) is not overly sensitive to the initialization of $\Aobjfixed$ and $\Aconfixed$, provided that a reasonable number of iterations are performed.
%
%
The results also suggest that Algorithm \ref{alg:outer_problem} without refinement (blue lines) requires an impractically large number of iterations to achieve comparatively small improvements of the lower bound.
%
Figure~\ref{fig:linquad_with_expanding_cx} (bottom) shows the cumulative computation time required for computing the approximate value functions, i.e., solving line~\ref{alg:inner_problem:lp} of Algorithm~\ref{alg:inner_problem} and line~\ref{alg:outer_problem:generate} of Algorithm~\ref{alg:outer_problem}.
%
This indicates that the hand-tuning method suggested in \cite{beuchat_2017_pwm_at_CDC} can achieve a tighter lower bound with less computation time compared to the method suggested in \cite{boyd_2013_iteratedApproxValueFunctions}. However, both methods have a limit beyond which extra computation did not improve the lower bound.
%
We note that the computation times for the results in Figure~\ref{fig:linquad} fall between the lines shown in Figure~\ref{fig:linquad_with_expanding_cx} (bottom), and are not shown for the sake of clarity.
%
The results on this higher dimensional example empirically support our claim that sub-gradients of the point-wise maximum objective function are effective for computing tighter lower bounds.

%% file: sec_for_arXiv_v2/06_conclusion.tex

\section{Conclusion} \label{sec:conclusion}

We proposed an algorithm that computes a family of lower bounding approximate value functions in an iterative fashion, with the choice of initial state distribution as the only parameter to be selected by the designer. We motivate our algorithm by considering the non-convex objective of maximizing the point-wise maximum of lower bounding value functions, and use sub-gradient information to find (potentially) sub-optimal solutions. Testing our algorithm on linear-quadratic examples, we demonstrated a significant tightening of the lower bound compared to existing methods, achieved with a modest or negligible increase in the computation time.

As future work, we will investigate adaptations of the proposed algorithm that are tailored to policy performance. This is a more challenging setting because the computation restriction are more stringent for evaluation of a policy. Moreover, counter-examples can readily be constructed where an approximate value function that provides a relatively tight lower bound leads to a greedy policy with relatively poor online performance.

A weakness of the proposed method is that by sampling from the initial state distribution we forgo the direct theoretical connection to the Bellman equation. An interesting direction for extending our algorithm is to consider methods that directly maximize the integral with respect to the initial state distribution, for example stochastic gradient ascent. This would also address the open question about whether the lower bound quality is sensitive to the choice of samples in our proposed algorithm.

%% file: sec_for_arXiv_v2/11_appendix_pwm_ineq_reformulation.tex

\section{Reformulation of Point-wise Maximum Inequality} \label{app:pwm_ineq_reformulation}

This appendix summarises our previous work \cite{beuchat_2017_pwm_at_CDC} in the context of this paper.

\subsection{Jensen's inequality and epigraph reformulation} \label{app:pwm_ineq_reformulation:jensen_epigraph}

The point-wise maximum constraint \eqref{eq:pwm_nlp_for_V:pwm_ineq} is equivalent to $J$ separate constraints of the form,
	\begin{equation} \nonumber
		\begin{aligned}
			l(x,u) \,+\, \disfac \, \expval{}{\max\limits_{k=1,\dots,J} \, V_k\left( g(x,u,\xi) \right)}
			\,.
		\end{aligned}
	\end{equation}
As $\max(\cdot)$ is a convex function, by Jensen's inequality a sufficient condition for constraint \eqref{eq:pwm_nlp_for_V:pwm_ineq} is,
	\begin{equation} \label{eq:app:pwm_ineq_reformulation:jensen}
		\begin{aligned}
			&V_j(x) \,\leq\, l(x,u)
				\,+\, \disfac \, \max\limits_{k=1,\dots,J} \, \expval{}{ V_k\left( g(x,u,\xi) \right) }
				\,,
			\\
			&\hspace{3.6cm} \forall \, \xinXcompact,\, \uinUcompact,\, j\!=\!1,\dots,J
				\,.
		\end{aligned}
	\end{equation}
%
An exact epigraph reformulation can now be applied \cite[Theorem 1]{beuchat_2017_ADPwithPenalty} with the epigraph variable denoted $s_V$, i.e.,
\begin{equation} \label{eq:pwm_ineq_sufficient}
	V_j(x) \,\leq\, l(x,u) \,+\, \gamma \, s_V^2
		\,,\hspace{0.1cm}
		\forall \, (x,u,s_V) \in \setepiV
		\,,
\end{equation}
for $j=1,\dots,J$, where the set $\setepiV$ is defined as,
\begin{equation} \nonumber
	\setepiV = \setdef{x,u,s_V}{
		\begin{matrix}
		\xinX,\, \uinU\,, s_V\in\rdim{},\,
		\\
		s_V^2 \geq \expval{}{ V_k\left( g(x,u,\xi) \right) } \; \forall \, k=1,\dots,J
		\end{matrix}
	}
	\,,
\end{equation}
We choose to square the epigraph variable $s_V$ without loss of generality because \cite[Assumptions 4.2.1(b)]{hernandez_2012_discreteTimeMCP} implies that $\Vopt$ is non-negative.

\subsection{S-procedure reformulation} \label{app:pwm_ineq_reformulation:sprocedure}

The S-procedure \cite{petersen_2012_robustHinfcontrol} is used to obtain a sufficient condition for \eqref{eq:pwm_ineq_sufficient}.
Applying the S-procedure to the relevant part of $\setepiV$ leads to,
	\begin{equation} \label{eq:app:pwm_ineq_reformulation:sprocedure}
		\begin{aligned}
			&V_j(x) \leq l(x,u) + \gamma \, s_V^2 - \sum\limits_{k=1}^J  \lambda_k \left( s_V^2 - \expval{}{ V_k\left( g(x,u,\xi) \right) } \right)
			,
			\\
			&\hspace{3.5cm} \forall \, \xinXcompact,\, \uinUcompact,\, s_V\!\in\!\rdim{},\, j\!=\!1,\dots,J
			,
		\end{aligned}
	\end{equation}
with $\lambda_k \!\in\! \rdim{}_+$ as the non-negative decision variables introduced by the S-procedure.
Reformulation \eqref{eq:app:pwm_ineq_reformulation:sprocedure} still suffers from difficulty (D\ref{difficulty:nccon}): there will be $J$ bilinear terms of the form $\smash{\lambda_k \, \expval{}{ V_k\left( g(x,u,\xi) \right) }}$ in each of the $J$ constraints.

The following implications summarize the approximation steps described,
\begin{equation} \nonumber
	\text{\eqref{eq:pwm_nlp_for_V:pwm_ineq}}
	\stackrel{\text{(Jensen)}}{\quad\arrL\quad}
	\text{\eqref{eq:app:pwm_ineq_reformulation:jensen}}
	\stackrel{\text{(Epigraph)}}{\quad\arrLR\quad}
	\text{\eqref{eq:pwm_ineq_sufficient}}
	\stackrel{\text{(S-procedure)}}{\quad\arrL\quad}
	\text{\eqref{eq:app:pwm_ineq_reformulation:sprocedure}}
	\,.
\end{equation}
%
In words, this reformulation is sufficient in the sense that if a family of functions $\smash{V_1,\dots, V_J}$ satisfies \eqref{eq:app:pwm_ineq_reformulation:sprocedure} then it also satisfies \eqref{eq:pwm_nlp_for_V:pwm_ineq} (but not necessarily the other way around).
%
An equivalent or tighter approximation can be found by allowing the S-procedure multipliers to depend on the state and input, i.e., $\smash{\lambda_k : \spaceX \!\times\! \spaceU \arrr \rdim{}_+}$. This would require the introduction of a restricted function space on $\smash{ \left(\spaceX\!\times\!\spaceU\right) }$, denoted $\approxFuncSpaceXU$, and defined similar to $\approxFuncSpaceX$ in \eqref{eq:approx_func_spaces}.

\subsection{Overcoming difficulty (D\ref{difficulty:nccon}) for \eqref{eq:canonical_inner_problem}} \label{app:pwm_ineq_reformulation:actual}

The auxiliary problem \eqref{eq:canonical_inner_problem} introduced in Section~\ref{sec:adp:pwm_objective} has a form similar to \eqref{eq:app:pwm_ineq_reformulation:jensen} except that the only constraint included is the one with the decision variable on the left side of the inequality. Letting $V_1$ in \eqref{eq:app:pwm_ineq_reformulation:jensen} correspond to $\alphaVbasis$ in \eqref{eq:canonical_inner_problem} and applying reformulation \eqref{eq:app:pwm_ineq_reformulation:sprocedure} we get the following sufficient condition for \eqref{eq:canonical_inner_problem:pwm_ineq},
	\begin{equation} \nonumber
		\begin{aligned}
			&\alphaVbasis(x) \,\leq\, l(x,u) \,+\, \gamma \, s_V^2
			\\
			&\hspace{1.6cm}
				-\, \sum\limits_{\alphaconfixed\in\Aconfixed}  \lambda_{\alphaconfixed} \left( s_V^2 - \alphaconfixed^\tran \expval{}{\Vbasis(g(x,u,\xi))} \right)
				,
		\end{aligned}
	\end{equation}
for all $\xinXcompact,\, \uinUcompact,\, s_V\!\in\!\rdim{}$, where the multipliers $\lambda_{\alphaconfixed}$ are additional non-negative decision variable for each element of $\Aconfixed$. As the $\alphaconfixed$ are fixed parameters in problem \eqref{eq:canonical_inner_problem}, it is clear that this reformulation is linear in the decision variables $\alpha$ and $\lambda_{\alphaconfixed}$. When the problem data and basis functions are polynomial, the infinite constraints are reformulated in the usual way, see \cite[Appendix A]{boyd_2014_iterated_bellman_inequality} for example, the result is a single Linear Matrix Inequality (LMI) constraint.

%% file: sec_for_arXiv_v2/11_appendix_inner_problem_proofs.tex

\section{Properties of the Inner Problem of Section \ref{sec:adp:gradient_algorithm}} \label{app:inner_problem}

All the material in this appendix is formulated for a minimization optimization objective, chosen to make the results readily comparable with existing optimization literature. Problem \eqref{eq:canonical_inner_problem} and Algorithm \ref{alg:inner_problem} are readily converted to minimization problems by taking the negative of the objective. 

\subsection{Differentiability definitions} \label{app:differentiability_definitions}

We provide for completeness the definitions of the \textbf{regular} and \textbf{general} sub-differential as taken from \cite[\S 7.D, \S8.A]{rockafellar_2009_variationalanalysis}.
%
The definition of the sub-differential commonly used for convex optimization problems is special case of the regular sub-differential defined below \cite[Proposition 8.12]{rockafellar_2009_variationalanalysis}, required here because \eqref{eq:canonical_inner_problem:objective} is non-convex when cast as a minimization problem. We require additionally the general sub-differential definition because \eqref{eq:canonical_inner_problem:objective} is non-smooth.

Given a function $\smash{f : \rdim{n} \arrr \rdim{} \cup \{-\infty,\infty\}}$, a vector $\smash{d\!\in\!\rdim{n}}$ is a \emph{regular lower subgradient} of $f$ at the point $\smash{x\!\in\!\rdim{n}}$ if the following one-sided limit condition holds,
%
	\begin{equation} \nonumber
		\liminf\limits_{z \arrr x,\, z\neq x}
			\hspace{0.1cm}
			\frac{f(z) - f(x) - \left(z-x\right)^\tran \, d }{ \left\| z - x \right\| }
			\,\geq\, 0
			\,.
	\end{equation}
The \emph{regular lower subdifferential} of $f$ at $x$, denoted $\smash{\regsubdiff f(x)}$ is the set of regular lower subgradients of $f$ at $x$.
%
A vector $\smash{d\!\in\!\rdim{n}}$ is a \emph{general lower subgradient} of $f$ at the point $x$ if there exists sequences $\smash{ x^{(i)} \!\stackrel{f}{\arrr}\! x }$ and $\smash{ d^{(i)} \!\arrr\! d }$ with $\smash{ d^{(i)} \!\in\! \regsubdiff f(x^{(i)}) }$, where the notation $\stackrel{f}{\arrr}$ stands for $f$-attentive, defined as,
	\begin{equation} \nonumber
		x^{(i)} \!\stackrel{f}{\arrr}\! x
			\hspace{0.4cm} \arrLR \hspace{0.4cm}
			x^{(i)} \!\arrr\! x
			\hspace{0.2cm} \text{with} \hspace{0.2cm}
			f(x^{(i)}) \!\arrr\! f(x)
			\,.
	\end{equation}
The \emph{general lower subdifferential} of $f$ at $x$, denoted $\smash{\gensubdiff f(x)}$ is the set of general lower subgradients of $f$ at $x$.
%
At a point $x$ where $f$ is finite, the set $\smash{\gensubdiff f(x)}$ and $\smash{\regsubdiff f(x)}$ are closed, with $\smash{\regsubdiff f(x)}$ convex and $\smash{\regsubdiff f(x) \!\subseteq\! \gensubdiff f(x)}$.
%
The function $f$ is \emph{subdifferentially regular} at a point $x$ if $\smash{\regsubdiff f(x) \!=\! \gensubdiff f(x)}$.
%
These definitions and properties correspond to \cite[Definition 8.3, Theorem 8.6, Definition 7.25]{rockafellar_2009_variationalanalysis}.
%
%
Note that if $f$ is differentiable at $x$, then $\smash{ \regsubdiff f(x) \!=\! \{ \nabla f(x) \} }$, i.e., a singleton, and if additionally $f$ is smooth on a neighbourhood of $x$, then $\smash{ \gensubdiff f(x) \!=\! \{ \nabla f(x) \} }$ also. For the standard definitions of the gradient $\nabla f(x)$ of a function $f$ at a differentiable point $x$, the reader is referred to \cite[\S B.5]{bertsekas_2016_nlprog_book}.
%
The regular and general \emph{upper} subdifferential are computed as $\smash{ -\regsubdiff\left(-f\right) }$ and $\smash{ -\gensubdiff\left(-f\right) }$, and denoted ${\regsubdiff^+ f}$ and ${\gensubdiff^+ f}$, respectively.

\subsection{Necessary condition for local optimality} \label{app:stationary_definition}

A function $\smash{f : \rdim{n} \arrr \rdim{}}$ is proper, for a minimisation objective, if $\smash{f(x)\!<\!+\infty}$ for at least one $\smash{x\!\in\!\rdim{n}}$, and $\smash{f(x)\!>\!-\infty}$ for all $\smash{x\!\in\!\rdim{n}}$.
%
Consider the minimization of a proper, lower-semi-continuous function $\smash{f : \rdim{n} \arrr \rdim{}}$ over a closed set $\constraintset \subseteq \rdim{n}$, i.e., $\smash{\min_{x\in\constraintset} f(x)}$. As per \cite[Theorem 8.15]{rockafellar_2009_variationalanalysis}, a necessary condition for the local optimality of a point $\overline{x} \in \constraintset$ is:
	\begin{equation} \label{eq:locally_optimal_definition}
		0 \,\in\, \gensubdiff f(\overline{x}) \,+\, \gennormalcone{\constraintset}(\overline{x})
			\,,
	\end{equation}
%
where $\gennormalcone{\constraintset}$ is the \emph{general normal cone} of the set $\constraintset$ at the point $\overline{x}$, see \cite[Definition 6.3]{rockafellar_2009_variationalanalysis}.
%
If in addition $\constraintset$ is a convex set, then this condition is equivalent to the existence of a $\smash{ d \!\in\! \gensubdiff f(\overline{x}) }$ satisfying
	\begin{equation} \label{eq:locally_optimal_definition_useful}
		\left(z-\overline{x}\right)^\tran \, d \,\geq\, 0
			\,,\hspace{0.3cm}
			\forall \, z \in \constraintset
			\,.
	\end{equation}
see \cite[Theorem 6.9]{rockafellar_2009_variationalanalysis}.
%
Note further that if $f$ is convex then these conditions are necessary and sufficient for $\overline{x}$ to be globally optimal.

A stationary point of the optimisation problem $\smash{\min_{x\in\constraintset} f(x)}$ is one satisfying $\smash{0 \!\in\! \gensubdiff\left(f(\overline{x})+\delta_\constraintset(\overline{x})\right)}$, where $\delta_\constraintset$ is the indicator function of the set $\constraintset$. All stationary points satisfy \eqref{eq:locally_optimal_definition} as,
	\begin{equation} \nonumber
		\gensubdiff\left(f(\overline{x}) + \delta_\constraintset(\overline{x})\right)
			\,\subseteq\,
			\gensubdiff f(\overline{x}) + \gensubdiff\delta_\constraintset(\overline{x})
			\,=\,
			\gensubdiff f(\overline{x}) + \gennormalcone{\constraintset}(\overline{x})
			\,.
	\end{equation}
If $\constraintset$ is convex, then the inclusion becomes an equality at a point $\overline{x}$ where $f$ is sub-differentially regular, \cite[Corollary 10.9]{rockafellar_2009_variationalanalysis}.

\subsection{Proof of convergence for a more general problem statement} \label{app:minimize_concave_function_convergence_proof}

To streamline the proof of Theorem \ref{thm:inner_problem_algorithm_convergence}, we consider now a more a general problem statement, and then in Appendix \ref{app:inner_problem_algorithm_convergence_proof} below we show that problem \eqref{eq:canonical_inner_problem} and Algorithm \ref{alg:inner_problem} has this form. Given a proper, lower-semi-continuous, \textbf{concave function} $\smash{f : \rdim{n} \arrr \rdim{}}$ and a \textbf{convex constraint} set $\constraintset \subseteq \rdim{n}$ such that $f$ is \textbf{bounded below} on $\constraintset$, we consider the optimization problem,
	\begin{equation} \label{eq:minimize_concave_function}
		\min \hspace{0.05cm} f(x)
			,\hspace{0.10cm}
			\subjto \hspace{0.1cm} x \in \constraintset
			\,.
	\end{equation}
We show that Algorithm \ref{alg:minimize_concave_function} finds points that satisfy \eqref{eq:locally_optimal_definition_useful}; note that in line \ref{alg:minimize_concave_function:subdiff} we use the lower sub-differential because \eqref{eq:minimize_concave_function} is a minimization problem.

\newcommand{\algthreeidx}{k}
%
\begin{figure}[h]
	\removelatexerror
	\begin{algorithm}[H]
		\caption{Find points satisfying necessary optimality conditions of problem \eqref{eq:minimize_concave_function}}
		\label{alg:minimize_concave_function}
		\begin{algorithmic}[1]
			%
			\Procedure{MinimizeConcaveFunction}{ $\smash{x^{(0)}}$ , $\epsilon$ }
			%
			\State $\algthreeidx\gets 0$
			%
			\Repeat
			%
			\State $d^{(\algthreeidx)} \gets$ an element from $\gensubdiff f\left( x^{(\algthreeidx)} \right)$
			\label{alg:minimize_concave_function:subdiff}
			%
			\If{$\left( d^{(\algthreeidx)} = 0 \right)$ }
			\label{alg:minimize_concave_function:zero_subdiff}
			%
			\State $x^{(\algthreeidx+1)} \gets x^{(\algthreeidx)}$
			\label{alg:minimize_concave_function:zero_subdiff:break}
			%
			\Else
			\State $x^{(\algthreeidx+1)} \gets \hspace{0.05cm} x^\ast \in \arg\min \left\{\, x^\tran \, d^{(\algthreeidx)} 	,\hspace{0.05cm} \subjto \hspace{0.05cm} x \in \constraintset \right\}$
			\label{alg:minimize_concave_function:lp}
			%
			\EndIf
			%
			\State $\algthreeidx\gets \algthreeidx+1$
			%
			\Until{$\Big( f\left(x^{(\algthreeidx)}\right) - f\left(x^{(\algthreeidx-1)}\right) \Big) < \epsilon$,}
			\label{alg:minimize_concave_function:equal_objective}
			%
			\State \textbf{return} $x^{(\algthreeidx)}$
			%
			\EndProcedure
		\end{algorithmic}
	\end{algorithm}
\end{figure}

\begin{theorem} \label{thm:minimize_concave_function_convergence}
	For any initial condition $\smash{x^{(0)} \!\in\! \constraintset}$ and any $\smash{\epsilon \!>\! 0}$, Algorithm \ref{alg:minimize_concave_function} generates a non-decreasing sequence $\smash{f\left(x^{(\algthreeidx)}\right)}$ and terminates after a finite number of iterations.
	%
	%
	%
	With $\smash{\epsilon \!=\! 0}$, and assuming that the $\arg\min$ on line~\ref{alg:minimize_concave_function:lp} is always attained, the sequence $\smash{f\left(x^{(\algthreeidx)}\right)}$, converges to a finite value, and the sequences $\smash{x^{(\algthreeidx)}}$, $\smash{d^{(\algthreeidx)}}$, satisfy condition~\eqref{eq:locally_optimal_definition_useful} in the following sense,
		\begin{equation} \nonumber
			\lim\limits_{\algthreeidx \arrr \infty}\left(\,
			\min\limits_{x\in\constraintset} \, \left(x-x^{(\algthreeidx)}\right)^\tran d^{(\algthreeidx)}
			\,\right)
			\,=\, 0
			\,.
		\end{equation}
\end{theorem}

\begin{IEEEproof}[Proof of Theorem \ref{thm:minimize_concave_function_convergence}]
	
	\textbf{The sub-differential gives majorizing functions}
	
	We first show that given any element of the general lower subdifferential, $\smash{d^{(\algthreeidx)} \!\in\! \gensubdiff f \left(x^{(\algthreeidx)}\right)}$, the surrogate function,
		\begin{equation} \label{eq:app:inner_problem_algorithm_convergence_proof:surrogate}
			s_{\algthreeidx}\left(x\right) \,=\,
				\left( x - x^{(\algthreeidx)} \right)^\tran \, d^{(\algthreeidx)}
				\,+\, f\left( x^{(\algthreeidx)} \right)
				,
		\end{equation}
	is a point-wise upper-bound of the concave function $f$. This is trivial for a differentiable point $\smash{x^{(\algthreeidx)}}$ as we have that the gradient is the only element of both the general lower and upper subdifferential of $f$ at $\smash{x^{(\algthreeidx)}}$ and hence $\smash{s_{\algthreeidx}\left(x\right)}$ is a global upper-bound of $f$. Pathological functions where the gradient is not an element of general subdifferential at a differentiable point are excluded by virtue of the $f$ being concave.

	At a non-differentiable point $\smash{x^{(\algthreeidx)}}$, as $f$ is concave, the regular lower subdifferential is empty at this point. Thus the general lower subdifferential is defined by the limits along all sequences of differentiable points leading to $\smash{x^{(\algthreeidx)}}$.
	As the regular lower and upper subdifferential are equal at all points along any such sequence we have that,
	\begin{equation} \nonumber
		\gensubdiff f (x^{(\algthreeidx)}) \,\subset\, -\gensubdiff\left( -f \right) (x^{(\algthreeidx)})
			\,.
	\end{equation}
	Thus is remains to show that $\gensubdiff\left( -f \right) (x^{(\algthreeidx)})$ contains only supporting hyperplanes of the hypograph of $f$ at $\smash{x^{(\algthreeidx)}}$. As $-f$ is convex, we have by \cite[Proposition 8.12]{rockafellar_2009_variationalanalysis} that the general lower subdifferential of $-f$ is,
		\begin{equation} \nonumber
			\begin{aligned}
				&\, \gensubdiff\left( -f(x^{(\algthreeidx)}) \right) 
					= \regsubdiff\left( -f(x^{(\algthreeidx)}) \right)
				\\
				=&\, \setdef{ -d \in \rdim{n} }{ f(x) \leq f(x^{(\algthreeidx)}) + \left(x-x^{(\algthreeidx)}\right)^\tran d ,\, \forall \xinXcompact }
				\,,
			\end{aligned}
		\end{equation}
	Thus we have shown that the surrogate function $\smash{ s_{\algthreeidx}\left(x\right) }$ is a point-wise upper-bound of the concave function $f$ at any point $\smash{x^{(\algthreeidx)}}$ that it is constructed.

	\vspace{0.2cm}
	\textbf{Termination in finite iterations}
	
	By definition, the minimization problem on line \ref{alg:minimize_concave_function:lp} returns $\smash{x^{(\algthreeidx+1)}}$ satisfying the optimality condition,
		\begin{equation} \label{eq:app:inner_problem_algorithm_convergence_proof:lp_optimality}
			\left(x-x^{(\algthreeidx+1)}\right)^\tran \, d^{(\algthreeidx)} \,\geq\, 0
				\,,\hspace{0.3cm}
				\forall \, x \in \constraintset
				\,.
		\end{equation}
	Combining the properties of the surrogate function $s_\algthreeidx$ with the definition of line \ref{alg:minimize_concave_function:lp} as a minimization problem, we have that,
		\begin{equation} \label{eq:app:inner_problem_algorithm_convergence_proof:nonincreasing}
			f\left( x^{(\algthreeidx)} \right)
				\,=\,
				s_{\algthreeidx}\left(x^{(\algthreeidx)}\right)
				\,\geq\,
				s_{\algthreeidx}\left(x^{(\algthreeidx+1)}\right)
				\,\geq\,
				f\left( x^{(\algthreeidx+1)} \right)
				,
				\hspace{-0.1cm}
		\end{equation}
	with $\smash{ x^{(\algthreeidx)},x^{(\algthreeidx+1)} \!\in\! \constraintset }$ ensured by the constraints of line \ref{alg:minimize_concave_function:lp}. The equality is by~\eqref{eq:app:inner_problem_algorithm_convergence_proof:surrogate}, the first inequality is by definition of the minimization on line \ref{alg:minimize_concave_function:lp}, and the final inequality is by the fact that the surrogate is a point-wise upper-bound.
	
	By the assumption that  $f$ is bounded below on $\constraintset$, the sequences $\smash{ f\left( x^{(\algthreeidx)} \right) }$ and $\smash{ s_{\algthreeidx}\left( x^{(\algthreeidx)} \right) }$, for $\smash{\algthreeidx\!\geq\!0}$, are convergent, hence Cauchy. Therefore, for all $\smash{\epsilon \!>\! 0}$ there must exist a $\smash{\algthreeidx \!\geq\! 1}$ such that the condition on line \ref{alg:inner_problem:equal_objective} triggers.

	\vspace{0.2cm}
	
	\textbf{Convergence to necessary conditions for optimality}
	
	For $\smash{\epsilon \!=\! 0}$ we have from the argument above that the sequences $\smash{ f\left( x^{(\algthreeidx)} \right) }$ and $\smash{ s_{\algthreeidx}\left( x^{(\algthreeidx)} \right) }$ converge to a finite value.
	%
	To show that the sequence $\smash{x^{(\algthreeidx)}}$ satisfies condition \eqref{eq:locally_optimal_definition_useful} in the limit, we need to show that,
		\begin{equation} \nonumber
			\lim\limits_{\algthreeidx \arrr \infty}\left(\,
				\sup\limits_{d\in\gensubdiff f (x^{(\algthreeidx)})} \left(\,
				\min\limits_{x\in\constraintset} \, \left(x-x^{(\algthreeidx)}\right)^\tran d
				\,\right)
				\,\right)
				\,\geq\, 0
				\,.
		\end{equation}
	%
	To show this it is sufficient to show that the sequence $\smash{ x^{(\algthreeidx)} }$ converges to an optimal point of $\smash{ \min\nolimits_{x\in\constraintset} s_\algthreeidx(x)}$, i.e., we show that sequences $\smash{ x^{(\algthreeidx)} }$, $\smash{ d^{(\algthreeidx)} }$, satisfy,
		\begin{equation} \label{eq:app:inner_problem_algorithm_convergence_proof:sufficient_limsup}
			\lim\limits_{\algthreeidx \arrr \infty}\left(\,
				\min\limits_{x\in\constraintset} \, \left(x-x^{(\algthreeidx)}\right)^\tran d^{(\algthreeidx)}
				\,\right)
				\,=\, 0
				\,.
		\end{equation}
	%
	The $\min$ here is attained by the assumption in the theorem statement that line~\ref{alg:minimize_concave_function:lp} of Algorithm~\ref{alg:minimize_concave_function} attains at every iteration.
	%
	To show that the limit in  \eqref{eq:app:inner_problem_algorithm_convergence_proof:sufficient_limsup} exists and equals zero, we first consider for the sake of contradiction that the sequences $\smash{x^{(\algthreeidx)}}$, $\smash{d^{(\algthreeidx)}}$ satisfy,
		\begin{equation} \nonumber
			\liminf\limits_{\algthreeidx \arrr \infty}\left(\,
				\min\limits_{x\in\constraintset} \, \left(x-x^{(\algthreeidx)}\right)^\tran d^{(\algthreeidx)}
				\,\right)
				\,=\, -\delta
				\,<\, 0
				\,.
		\end{equation}
	By definition of the $\liminf$, for every $\smash{\algthreeidx\!\geq\!0}$ there exists a $\smash{j\!\geq\!\algthreeidx}$ for which,
		\begin{equation} \label{eq:app:inner_problem_algorithm_convergence_proof:contradiction_limsup}
			\min\limits_{x\in\constraintset} \, \left(x-x^{(j)}\right)^\tran d^{(j)}
				\,\leq\, - \,\frac{\delta}{2}
				\,.
		\end{equation}
	By definition of line \ref{alg:minimize_concave_function:lp} as a minimization problem we have for this pair $\algthreeidx,j$ that,
	\begin{equation} \nonumber
		\begin{aligned}
			s_{j+1}\left( x^{(j+1)} \right)
			\,&\stackrel{\eqref{eq:app:inner_problem_algorithm_convergence_proof:nonincreasing}}{\leq}\,
			s_{j}\left( x^{(j+1)} \right)
			\\
			&\stackrel{\eqref{eq:app:inner_problem_algorithm_convergence_proof:surrogate}}{=}\,
			\left( x^{(j+1)} - x^{(j)} \right)^\tran \, d^{(j)} \,+\, f\left( x^{(j)} \right)
			\\
			&\stackrel{\eqref{eq:app:inner_problem_algorithm_convergence_proof:contradiction_limsup}}{\leq}\,
			-\,\frac{\delta}{2} \,+\, f\left( x^{(j)} \right)
			\\
			\,&\stackrel{\eqref{eq:app:inner_problem_algorithm_convergence_proof:nonincreasing}}{\leq}\,
			-\,\frac{\delta}{2} \,+\, s_{\algthreeidx}\left( x^{(\algthreeidx)} \right)
			\,.
		\end{aligned}
	\end{equation}
	Repeating this argument starting from $j+1$, we readily establish that,
		\begin{equation} \nonumber
			\limsup\limits_{\algthreeidx\arrr\infty} \, s_{\algthreeidx}\left( x^{(\algthreeidx)} \right)
				\,\leq\, \limsup\limits_{N\arrr\infty} \left(\,
				s_{0}\left(x^{(0)}\right) \,-\, N \, \frac{\delta}{2}
				\,\right)
				\,=\, -\infty
				\,,
		\end{equation}
	which contradicts the previous conclusion that the sequence $\smash{ s_{\algthreeidx}\left( x^{(\algthreeidx)} \right) }$ converges to a finite value.
	Moreover we have that,
		\begin{equation} \nonumber
			\min\limits_{x\in\constraintset} \, \left(x-x^{(\algthreeidx)}\right)^\tran d^{(\algthreeidx)}
			\,\leq\, 0
			\,,\hspace{0.3cm}
			\text{for $\smash{\algthreeidx\!\geq\!0}$,}
		\end{equation}
	because $\smash{x^{(\algthreeidx)} \!\in\! \constraintset}$, for $\smash{\algthreeidx\!\geq\!0}$.
	%
	Thus, by contradiction we have shown that,
		\begin{equation} \nonumber
			\begin{aligned}
				0 \,\geq&\, \limsup\limits_{\algthreeidx \arrr \infty}\left(\,
				\min\limits_{x\in\constraintset} \, \left(x-x^{(\algthreeidx)}\right)^\tran d^{(\algthreeidx)}
				\,\right)
				\\
				\,\geq&\,\,
				\liminf\limits_{\algthreeidx \arrr \infty}\left(\,
				\min\limits_{x\in\constraintset} \, \left(x-x^{(\algthreeidx)}\right)^\tran d^{(\algthreeidx)}
				\,\right)
				\,\geq\, 0
				\,,
			\end{aligned}
		\end{equation}
	and hence the limit in~\eqref{eq:app:inner_problem_algorithm_convergence_proof:sufficient_limsup} exists and equals zero.
\end{IEEEproof}

\vspace{0.1cm}

Note that if line \ref{alg:minimize_concave_function:zero_subdiff} of Algorithm \ref{alg:minimize_concave_function} triggers, then the subgradient is zero and condition \eqref{eq:locally_optimal_definition_useful} is satisfied. In this case the $x^{(\algthreeidx)}$ returned is a global maximizer of the concave function $f$.
%
Note also that for a positive $\epsilon$, if the condition on line \ref{alg:inner_problem:equal_objective} triggers with $\smash{ f\left(x^{(\algthreeidx)}\right) = f\left(x^{(\algthreeidx-1)}\right) }$, then $x^{(\algthreeidx-1)}$ satisfies \eqref{eq:locally_optimal_definition_useful}.
To show this, first note that by \eqref{eq:app:inner_problem_algorithm_convergence_proof:surrogate} and \eqref{eq:app:inner_problem_algorithm_convergence_proof:nonincreasing} we have,
\begin{equation} \nonumber
	\begin{aligned}
		f\left( x^{(\algthreeidx-1)} \right)
		&\,\stackrel{\eqref{eq:app:inner_problem_algorithm_convergence_proof:nonincreasing}}{=}\,
		s_{\algthreeidx-1}\left(x^{(\algthreeidx)}\right)
		\\
		&\,\stackrel{\eqref{eq:app:inner_problem_algorithm_convergence_proof:surrogate}}{=}\,
		\left( x^{(\algthreeidx)} - x^{(\algthreeidx-1)} \right)^\tran  d^{(\algthreeidx-1)}
		+ f\left( x^{(\algthreeidx-1)} \right)
		.
	\end{aligned}
\end{equation}
From this we substitute $\smash{x^{(\algthreeidx)\tran}  d^{(\algthreeidx-1)} = x^{(\algthreeidx-1)\tran} d^{(\algthreeidx-1)} }$ into the optimality condition \eqref{eq:app:inner_problem_algorithm_convergence_proof:lp_optimality} that  $\smash{x^{(\algthreeidx)}}$ satisfies, and we get that the $\smash{d^{(\algthreeidx-1)} \!\in\! \gensubdiff f \left(x^{(\algthreeidx-1)}\right)}$ from line \ref{alg:minimize_concave_function:subdiff} of Algorithm \ref{alg:minimize_concave_function} satisfies condition \eqref{eq:locally_optimal_definition_useful} at $\smash{x^{(\algthreeidx-1)}}$.

\subsection{Proof of convergence for Algorithm \ref{alg:inner_problem}} \label{app:inner_problem_algorithm_convergence_proof}

\begin{IEEEproof}[Proof of Theorem \ref{thm:inner_problem_algorithm_convergence}]
	
	We show that the objective function $\pwmobjfuncsymbol$ and the convex constraint $\smash{\alpha \!\in\! \BIsufficient(\Aconfixed)}$ satisfy the assumptions of Theorem \ref{thm:minimize_concave_function_convergence}. Casting~\eqref{eq:canonical_inner_problem} as a minimization problem, the objective is,
		\begin{equation} \nonumber
			-\pwmobjfuncsymbol\left( \alpha \right) =
				-\sum\limits_{i=1}^{N_c} \, \left(\, \max\left\{ \alphaVbasis(x_{c,i}) , \Vobjfixed(x_{c,i}) \right\} \,\right)
				\,.
		\end{equation}
	The two elements of the $\max$ are linear in the decision variable $\alpha$, and thus the objective is concave in $\alpha$.

	We now show that $-\pwmobjfuncsymbol$ is bounded below on the constraint set. The assumption that $\Vconfixed$ is a point-wise lower bound of $\Vopt$ means that for any $\alpha$ satisfying constraint \eqref{eq:canonical_inner_problem:pwm_ineq}, the function $\smash{ \alphaVbasis(x) }$ is also a point-wise lower bound of $\smash{\Vopt(x)}$ for all $\xinXcompact$. This is ensured by the Bellman operator $\bellmanOp[]{}$ being monotone and $\disfac$-contractive. Moreover, under \cite[Assumptions 4.2.1(a), 4.2.1(b), 4.2.2]{hernandez_2012_discreteTimeMCP} we have that $\smash{\Vopt(x)}$ is finite for all $\xinXcompact$. As $\Vobjfixed$ is also a point-wise lower bound of $\Vopt$, we have that all elements of the sum in $\pwmobjfuncsymbol$ are bounded above, and hence $-\pwmobjfuncsymbol$ is bounded below for all $\smash{\alpha \!\in\! \BIsufficient(\Aconfixed)}$.

	Next we show that equation \eqref{eq:pwmobj_subdiff} correctly computes an element of the general upper subdifferential of $\pwmobjfuncsymbol$. For an $\alpha$ where $\pwmobjfuncsymbol$ is differentiable, we have that $\smash{ \alphaVbasis(x_{c,i}) \!\neq\! \Vobjfixed(x_{c,i}) }$ for all $\smash{i=1,\dots,N_c}$, and thus equation \eqref{eq:pwmobj_subdiff} computes the gradient at this point. The objective function $\pwmobjfuncsymbol$ is non-differentiable for an $\alpha$ where $\smash{ \alphaVbasis(x_{c,i}) \!=\! \Vobjfixed(x_{c,i}) }$ for at least one point $\smash{i=1,\dots,N_c}$. Letting $\mcal{I}_{<}(\alpha)$, $\mcal{I}_{=}(\alpha)$, and $\mcal{I}_{>}(\alpha)$ denote the indices $\smash{i=1,\dots,N_c}$ where $\smash{\alphaVbasis(x_{c,i})}$ is respectively less than, equal, and greater than $\smash{\Vobjfixed(x_{c,i}) }$, we define $\delta_{\min}(\alpha)$ as,
		\begin{equation} \nonumber
				\delta_{\min}(\alpha) \,=\,
					\min_{i \,\in\, \big(\, \mcal{I}_{<}(\alpha) \,\cup\, \mcal{I}_{>}(\alpha) \,\big) }
					\hspace{0.1cm}
					\left| \alphaVbasis(x_{c,i}) - \Vobjfixed(x_{c,i}) \right|
					\,.
		\end{equation}
	Recall that under Assumption \ref{ass:basisfunctions}, $\Vbasis_1$ is taken to be the constant function and let $e_1$ denote a vector with $1$ as the first element and zero otherwise. Thus for all $\smash{ \delta \!\in\! (0,\delta_{\min}) }$ we have that $\pwmobjfuncsymbol$ is differentiable at $\smash{ \left(\alpha+\delta \, e_1 \right) }$ with gradient given by equation \eqref{eq:pwmobj_subdiff}.
	%
	For any sequence $\delta \arrr 0$, the sequence $\smash{ \left(\alpha+\delta \, e_1 \right) }$ is $\pwmobjfuncsymbol$-attentive, i.e., $\smash{\pwmobjfuncsymbol\left(\alpha+\delta \, e_1 \right) \arrr \pwmobjfuncsymbol\left(\alpha\right)}$ by continuity of $\pwmobjfuncsymbol$.
	%
	As the gradient is the same for all $\smash{ \delta \!\in\! (0,\delta_{\min}) }$, equation \eqref{eq:pwmobj_subdiff} correctly computes an element of the general upper subdifferential of $\pwmobjfuncsymbol$ at $\alpha$.

	Finally, we need to show that the maximum on line \ref{alg:inner_problem:lp} of Algorithm \ref{alg:inner_problem} is always attained. First note that the objective coefficient vector on line \ref{alg:inner_problem:lp} of Algorithm \ref{alg:inner_problem} is given by,
		\begin{equation} \nonumber
			d^{(i)} \,=\, \sum\nolimits_{\big(\mcal{I}_{>}(\alpha^{(i)}) \,\cup\, \mcal{I}_{=}(\alpha^{(i)})\big)} \, \Vbasis(x_{c,i})
				\,.
		\end{equation}
	%
	By Assumption \ref{ass:lp_attainment} we have that,
		\begin{equation} \nonumber
			\max_{\alpha\in\rdim{K}}\left\{ \alpha^\tran \Vbasis(x_{c,i}) \,;\, \subjto\, \alpha \in \BIsufficient(\Aconfixed) \right\}
		\end{equation}
	attains its maximum for all $\smash{x_{c,i}}$, $\smash{i\!=\!1,\dots,N_c}$, and denote $f_i^\ast$ as the optimal value.
	%
	Thus the hyperplanes $\smash{\alpha^\tran \Vbasis(x_{c,i}) \!\leq\! f_i^\ast}$ are all supporting hyperplanes of the convex constraint set $\BIsufficient(\Aconfixed)$. The following finite dimensional linear program relaxation of line \ref{alg:inner_problem:lp} of Algorithm \ref{alg:inner_problem} also attains its maximum,
		\begin{equation} \label{eq:inner_problem_algorithm_convergence:proof:finiteLP}
			\begin{aligned}
				\max\limits_{\alpha \in \rdim{K}}
					\hspace{0.2cm}
					& \sum\nolimits_{\left(\mcal{I}_{>} \,\cup\, \mcal{I}_{=}\right)} \, \alpha^\tran \, \Vbasis(x_{c,i})
				\\
				\subjto \hspace{0.2cm}
					& \alpha^\tran \Vbasis(x_{c,i}) \leq f_i^\ast
					\,,\hspace{0.1cm} i=1,\dots,N_c
					\,.
			\end{aligned}
		\end{equation}
	%
	To show this, first observe that~\eqref{eq:inner_problem_algorithm_convergence:proof:finiteLP} is feasible and bounded above by $\sum\nolimits_{\left(\mcal{I}_{>} \,\cup\, \mcal{I}_{=}\right)} \smash{f_i^\ast}$, and thus by \cite[Corollary 27.3.2]{rockafellar_2015_convexanalysis} problem~\eqref{eq:inner_problem_algorithm_convergence:proof:finiteLP} attains its maximum. 
	%
	Finally, by \cite[Corollary 27.3.3]{rockafellar_2015_convexanalysis} we have that attainment for \eqref{eq:inner_problem_algorithm_convergence:proof:finiteLP} implies attainment for line \ref{alg:inner_problem:lp} of Algorithm \ref{alg:inner_problem}.
	
	We have shown that the assumptions of Theorem \ref{thm:inner_problem_algorithm_convergence} satisfy also the assumptions of Theorem \ref{thm:minimize_concave_function_convergence} and hence the claims follow from Theorem \ref{thm:minimize_concave_function_convergence}.
\end{IEEEproof}

%% file: sec_for_arXiv_v2/11_appendix_formulation_for_solver.tex

\section{Problem setting for Section \ref{sec:numerical}} \label{app:implementation_details}

\subsection{Quadratic Basis Functions} \label{app:quadratic_basis_functions}

The space of quadratic functions is parameterized by a constant offset $\smash{s\in\rdim{}}$, a linear co-efficient $\smash{p\in\rdim{n_x}}$, and a quadratic coefficient as a symmetric matrix $\smash{P\in\sdim{n_x}}$. Thus we express the restricted function space as,
\begin{equation} \label{eq:quadratic_basis_function_space}
	\begin{aligned}
		\approxFuncSpaceX =
		\setdef{ \hat{V}(x) }{
			\begin{matrix}
			V(x) \!=\! x^\tran P x + p^{\tran} x + s^{}
			\\
			P \in \sdim{n_x} \,,\,\, p \in \rdim{n_x} \,,\,\, s \in \rdim{}
			\end{matrix}
		}
		\,.
	\end{aligned}
\end{equation}
Thus the $\alpha$ is the stacked vector of $s$, $p$, and the unique elements of $P$, and the basis functions $\Vbasis$ are the monomials of $x$ up to degree two, which clearly satisfy Assumption \ref{ass:basisfunctions}.
%
Similar to Section \ref{sec:adp:existing} we use a subscript on $s$, $p$, and $P$ to label the approximate value function they correspond to, for example, $\smash{\Vhat_j\!\in\!\approxFuncSpaceX}$ is equivalent to $\smash{\Vhat_j(x) \!=\! x^\tran P_j x + p_j^{\tran} x + s_j}$ for all $\xinXcompact$.
%
This space of convex quadratic functions is considered by restricting matrix $P$ to be positive semi-definite.

\subsection{Formulating line \ref{alg:inner_problem:lp} of Algorithm \ref{alg:inner_problem} for commercial solver} \label{app:solver_formulation}

See Section \ref{sec:numerical} for the definitions of $A$, $B_u$, and $B_\xi$ as the linear dynamics, and Appendix \ref{app:quadratic_basis_functions} for the specification of the quadratic basis functions.
%
We introduce $\smash{\underline{u}_i,\overline{u}_i \in \mbb{R}}$, $\smash{i\!=\!1,\dots,n_u}$, with $\smash{\underline{u}_i < \overline{u}_i}$, to denote the lower and upper bounds that describe each coordinate of the $\smash{\mcal{U} \subseteq \mbb{R}^{n_u}}$ space.
%
The quadratic stage cost is condensed into the matrix $\smash{L \in \mbb{R}^{(n_x+n_u+1)\times(n_x+n_u+1)}}$ that takes the form $\smash{l(x,u) = [x^\tran,u^\tran,1] \, L \, [x^\tran,u^\tran,1]^\tran}$.
%
The notation $\smash{\diag{\cdot}}$ places the vector argument on the diagonal of an otherwise zero matrix, and $e_i$ is the standard basis column vector with $1$ in the $i^{\mrm{th}}$ element and zeros elsewhere, with the dimension clear from context.
%
We overload the notation $\Vhat$ and introduce the notation $\hat{\EVof}$ as the following matrices,
\begin{subequations} \nonumber
	\begin{align}
		\Vhat \,&=\,
		\begin{bmatrix}
		P & 0 & \frac{1}{2} p
		\\
		\star & 0 & 0
		\\
		\star & \star & s
		\end{bmatrix}
		\,,
		\\
		\hat{\EVof} \,&=\,
		\begin{bmatrix}
		A^\tran P A
		& A^\tran P B_u
		& \frac{1}{2} A^\tran p + A^\tran P B_\xi \expval{}{\xi}
		\\
		\star
		& B_u^\tran P B_u
		& \frac{1}{2} B_u^\tran p + B_u^\tran P B_\xi \expval{}{\xi}
		\\
		\star
		& \star
		& s + \trace{ B_\xi^\tran P B_\xi \expval{}{\xi \xi^\tran}}
		\end{bmatrix}
		\,,
	\end{align}
\end{subequations}
where $\star$ indicates that the matrix is symmetric. Again, any subscript $\Vhat_{(\cdot)}$, $\EVof_{(\cdot)}$ also applies to $s$, $p$, and $P$. Both matrices are symmetric with dimension $\smash{(n_x+n_u+1)}$.

Using this notation, the point-wise maximum Bellman inequality \eqref{eq:canonical_inner_problem:pwm_ineq}, repeated here for convenience,
\begin{equation} \nonumber
	\alphaVbasis(x) \leq \left( \bellmanOp[u]{} \Vconfixed \right) (x,u)
		\,, \hspace{0.2cm}
		\forall \, \xinXcompact ,\, \uinUcompact
		\,,
\end{equation}
is sufficiently reformulated as the following LMI:
\begin{equation} \label{eq:app:solver_formulation:LMI}
\begin{aligned}
	\hspace{-0.10cm}
	0 \preceq&
		\,-\,
		\begin{bmatrix}
		\Vhat & 0
		\\
		\star & 0
		\end{bmatrix}
		\,+\,
		\begin{bmatrix} L & 0 \\ \star & 0 \end{bmatrix}
		\,+\, 
		\begin{bmatrix}
		0 & 0
		\\
		\star & \gamma
		\end{bmatrix}
	\\
	&\,-\, \sum\limits_{\bar{\alpha}\in\Aconfixed} \, \lambda_{\bar{\alpha}} \,
	\begin{bmatrix}
	-\hat{\EVof}_{\bar{\alpha}} & 0
	\\
	\star & 1
	\end{bmatrix}
	\\
	&\,-\, \sum\limits_{i=1}^{n_u} \, \lambda_i \,
	\begin{bmatrix}
	0_{n_x \times n_x} & 0 & 0
	\\
	\star & -\smash{\diag{e_i}}
	& \smash{\frac{1}{2}\left( \underline{u}_i + \overline{u}_i \right)} e_i
	\\
	\star
	& \star
	& -\underline{u}_i \, \overline{u}_i
	\end{bmatrix}
	\!.
\end{aligned}
\end{equation}
%
The $s$, $p$, and $P$ in $\Vhat$ are decision variables, as well as the $\smash{\lambda_i \in \mbb{R}_+}$ and $\smash{\lambda_{\bar{\alpha}} \in \mbb{R}_+}$, with everything else as fixed problem data.
The $\smash{\lambda_i}$ are the auxiliary variables introduced when using the S-procedure to reformulate the for all $\smash{\uinU}$ part of the constraint in Appendix \ref{app:pwm_ineq_reformulation:actual}, while the $\smash{\lambda_{\bar{\alpha}}}$ are the auxiliary variables described in Appendix \ref{app:pwm_ineq_reformulation:actual}.
%
The objective function on line~\ref{alg:inner_problem:lp} of Algorithm~\ref{alg:inner_problem} is linear in the decision variables, and when computed as per line~\ref{alg:inner_problem:subdiff} of Algorithm~\ref{alg:inner_problem} it requires computation of the first and second moments of the $x_{c,i}$ for the indices, $\smash{i=1,\dots,N_c}$, where the approximate value function under consideration dominates $\Vconfixed$. Letting $\mu_c$ and $\Sigma_c$ denote the first and second moments respectively, the problem on line~\ref{alg:inner_problem:lp} of Algorithm~\ref{alg:inner_problem} becomes
\begin{equation} \label{eq:adp:lp_approach_with_LMI}
	\max\limits_{s,p,P} \, \left\{\, \trace{P \Sigma_c} + p^\tran \mu_c + s ,\; \subjto \text{\eqref{eq:app:solver_formulation:LMI}} \,\right\}
	\,,
\end{equation}
where $\trace{\cdot}$ denotes the trace of a square matrix.
%
Note that the constraint $\smash{P\,\succeq\,0}$ can be added to restrict to the space of convex quadratic functions.

%% file: sec_for_arXiv_v2/99_biography.tex
%
%

\begin{IEEEbiography}[{\includegraphics[width=1in,height=1.25in,clip,keepaspectratio]{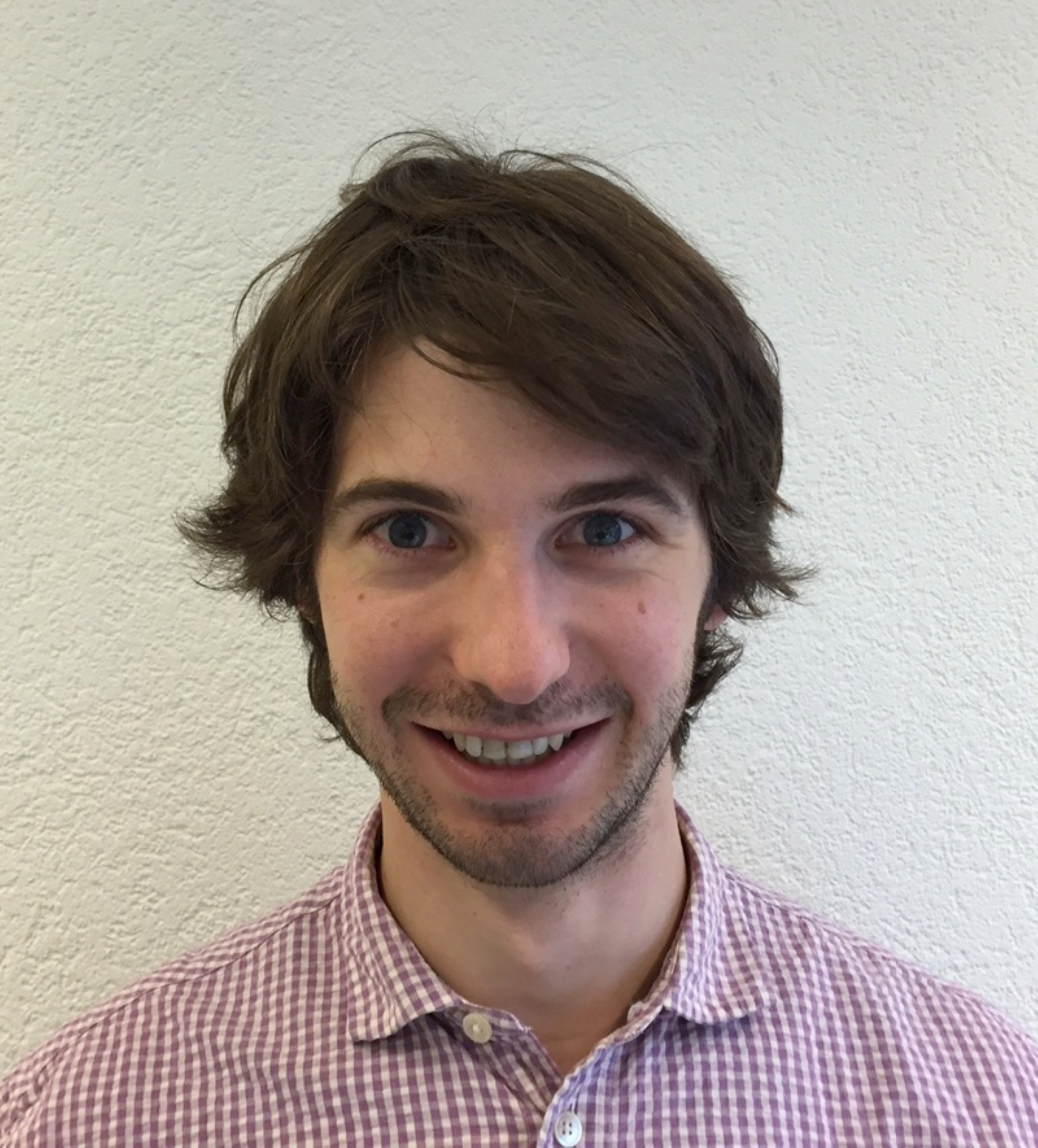}}]{Paul N. Beuchat}
	received the B.Eng. degree in mechanical engineering and B.Sc. in physics from the University of Melbourne, Australia, in 2008, and the M.Sc. degree in robotics, systems and control from ETH Z\"{u}rich, Switzerland, in 2014, where he is currently working towards the Ph.D degree at the Automatic Control Laboratory.
	From 2009-2012 he was as a subsurface engineer for ExxonMobil.
	His research interests are control and optimization of large scale systems, with a focus towards developing approximate dynamic programming techniques for applications in the areas of building control, and coordinated flight.
\end{IEEEbiography}

\begin{IEEEbiography}[{\includegraphics[width=1in,height=1.25in,clip,keepaspectratio]{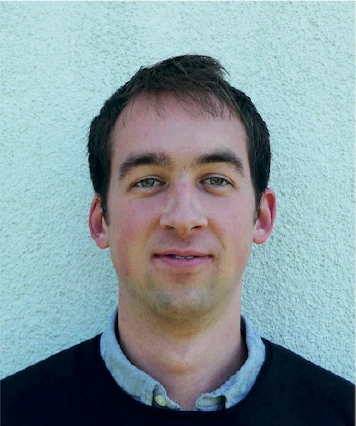}}]{Joseph Warrington}
	is a Senior Scientist in the Automatic Control Lab (IfA) at ETH Zurich, Switzerland. His Ph.D.~is from ETH Zurich (2013), and his B.A.~and M.Eng.~degrees in Mechanical Engineering are from the University of Cambridge (2008). From 2014-2016 he worked as an energy consultant at Baringa Partners LLP, London, UK, and he has also worked as a control systems engineer at Wind Technologies Ltd., Cambridge, UK, and privately as an operations research consultant. He is the recipient of the 2015 ABB Research Prize for an outstanding PhD thesis in automation and control, and a Simons-Berkeley Fellowship for the period January-May 2018. His research interests include dynamic programming, large-scale optimization, and predictive control, with applications including power systems and transportation networks.
\end{IEEEbiography}

\begin{IEEEbiography}[{\includegraphics[width=1in,height=1.25in,clip,keepaspectratio]{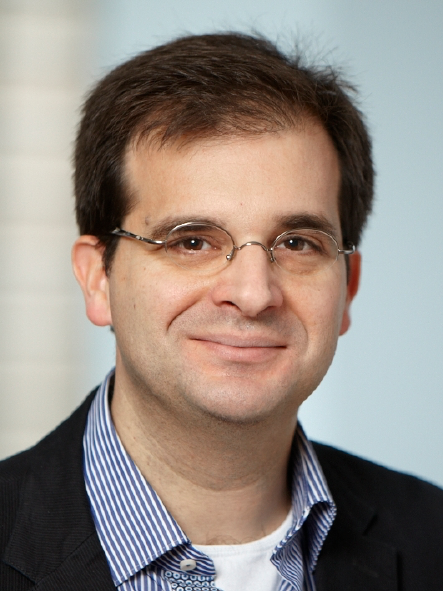}}]{John Lygeros}
	completed a B.Eng. degree in electrical engineering in  1990 and an M.Sc. degree in Systems Control in 1991, both at Imperial College of Science Technology and Medicine, London, U.K.. In 1996 he obtained a Ph.D. degree from the Electrical Engineering and Computer Sciences Department, University of California, Berkeley. During the period 1996--2000 he held a series of post-doctoral researcher appointments at the Laboratory for Computer Science, M.I.T., and the Electrical Engineering and Computer Sciences Department at U.C. Berkeley. Between 2000 and 2003 he was a University Lecturer at the Department of Engineering, University of Cambridge, U.K., and a Fellow of Churchill College. Between 2003 and 2006 he was an Assistant Professor at the Department of Electrical and Computer Engineering, University of Patras, Greece. In July 2006 he joined the Automatic Control Laboratory at ETH Zurich, where he is currently serving as the Head of the Automatic Control Laboratory and the Head of the Department of Information Technology and Electrical Engineering. His research interests include modelling, analysis, and control of hierarchical, hybrid, and stochastic systems, with applications to biochemical networks, automated highway systems, air traffic management, power grids and camera networks. John Lygeros is a Fellow of the IEEE, and a member of the IET and the Technical Chamber of Greece; since 2013 he serves as the Treasurer of the International Federation of Automatic Control.
\end{IEEEbiography}

\vfill





